\documentclass[manuscript]{aastex6}   		
\usepackage{graphicx}
\usepackage{subfloat}
\usepackage{amssymb}
\usepackage{amsmath}
\usepackage{wasysym}
\usepackage{hyperref}
\usepackage{epsfig, natbib}
\usepackage{color,soul}

\newcommand{\RNum}[1]{\uppercase\expandafter{\romannumeral #1\relax}}
\usepackage{afterpage}
\newcommand{\Hi}{H~{\sc i}}
\newcommand{\Hii}{H~{\sc ii}}

\shorttitle{G23.33-0.30}
\shortauthors{Hogge et al.}

\begin{document}

\title{The Interaction Between the Supernova Remnant W41 and the Filamentary Infrared Dark Cloud G23.33-0.30}
\author{Taylor G. Hogge\altaffilmark{1}, James M. Jackson\altaffilmark{2,3,1}, David Allingham\altaffilmark{3}, Andres E. Guzman\altaffilmark{4}, Nicholas Killerby-Smith\altaffilmark{3}, Kathleen E. Kraemer\altaffilmark{6}, Patricio Sanhueza\altaffilmark{4}, Ian W. Stephens\altaffilmark{5}, J. Scott Whitaker\altaffilmark{7}}

\altaffiltext{1}{\itshape Institute for Astrophysical Research, 725 Commonwealth Ave., Boston University, Boston, MA 02215, USA; thogge@bu.edu}
\altaffiltext{2}{\itshape SOFIA Science Center, USRA, NASA Ames Research Center, Moffett Field CA 94045, USA}
\altaffiltext{3}{\itshape School of Mathematical and Physical Sciences, University of Newcastle, University Drive, Callaghan NSW 2308, Australia}
\altaffiltext{4}{\itshape National Astronomical Observatory of Japan, National Institutes of Natural Sciences, 2-21-1 Osawa, Mitaka, Tokyo 181-8588, Japan}
\altaffiltext{5}{\itshape Harvard-Smithsonian Center for Astrophysics, 60 Garden Street, Cambridge, MA 02138, USA}
\altaffiltext{6}{\itshape Institute for Scientific Research, Boston College, 140 Commonwealth Avenue, Chestnut Hill, MA 02467, USA}
\altaffiltext{7}{\itshape Physics Department, 590 Commonwealth Ave., Boston University, Boston, MA 02215, USA}

\begin{abstract}

G23.33-0.30 is a 600~$M_{\odot}$ infrared dark molecular filament that exhibits large NH$_3$ velocity dispersions ($\sigma \sim 8 \ \rm{km \ s^{-1}}$) and bright, narrow NH$_3$(3,3) line emission. We have probed G23.33-0.30 at the $<0.1$ pc scale and confirmed that the narrow NH$_3$(3,3) line is emitted by four rare NH$_3$(3,3) masers, which are excited by a large-scale shock impacting the filament. G23.33-0.30 also displays a velocity gradient along its length, a velocity discontinuity across its width, shock-tracing SiO(5-4) emission extended throughout the filament, broad turbulent line widths in NH$_3$(1,1) through (6,6), CS(5-4), and SiO(5-4), as well as an increased NH$_3$ rotational temperature ($T_{\rm{rot}}$) and velocity dispersion ($\sigma$) associated with the shocked, blueshifted component. The correlations among $T_{\rm{rot}}$, $\sigma$, and $V_{\rm{LSR}}$ implies that the shock is accelerating, heating, and adding turbulent energy to the filament gas. Given G23.33-0.30's location within the giant molecular cloud G23.0-0.4, we speculate that the shock and NH$_3$(3,3) masers originated from the supernova remnant W41, which exhibits additional evidence of an interaction with G23.0-0.4. We have also detected the 1.3 mm dust continuum emission from at least three embedded molecular cores associated with G23.33-0.30. Although the cores have moderate gas masses ($M = 7-10$ M$_{\odot}$), their large virial parameters ($\alpha=4-9$) suggest that they will not collapse to form stars. The turbulent line widths of the cores may indicate negative feedback due to the SNR shock.

\end{abstract}

\section{Introduction}
\label{sec:intro}

High-mass stars ($M>8 \ \rm{M_{\odot}}$), though rare, have a profound impact on the evolution of the interstellar medium (ISM). Throughout their short lifetimes ($\sim 10^6$ yr), high-mass stars will release fast, radiation-driven stellar winds that carve out \Hii\ regions into the surrounding molecular clouds (MCs). High-mass stars end their lives by releasing $\sim 10^{51}$ ergs of energy nearly instantaneously in the form of supernovae (SNe). Shocks from expanding \Hii\ regions and supernova remnants (SNRs) accelerate, heat, and add turbulence to their surrounding gas. While these feedback mechanisms are thought to play a significant role in regulating the star formation process, the exact role these shocks play is not yet well understood. The results from hydrodynamic simulations of SNR-MC interactions differ depending on specific parameters, but they show that such interactions could potentially aid or inhibit star formation. Some models suggest that shocks act to disperse dense gas structures \citep{2016MNRAS.457.4470P}. On the other hand, since post-shock gas is compressed, other models suggest that this shock-induced compression can trigger collapse to form even denser compact structures \citep{2005A&A...444..505O}. Because the influence of SNe on the star formation process has implications for galaxy evolution, it is important to gain a better understanding of SNR-MC interactions. The sites of these interactions can be associated with broad molecular line widths \citep{2005ApJ...618..297R} and shock-excited maser emission in the 1720 MHz OH transition \citep{2002Sci...296.2350W} and the NH$_3$(3,3) inversion transition \citep{2016ApJ...826..189M}. Dense molecular gas structures that exhibit evidence of broad molecular line widths, 1720 MHz OH maser emission, or NH3 (3,3) maser emission, but lack an obvious stellar source, are prime targets to investigate potential SNR-MC interactions.

\begin{figure}[h]
\centering
\includegraphics[scale=0.8]{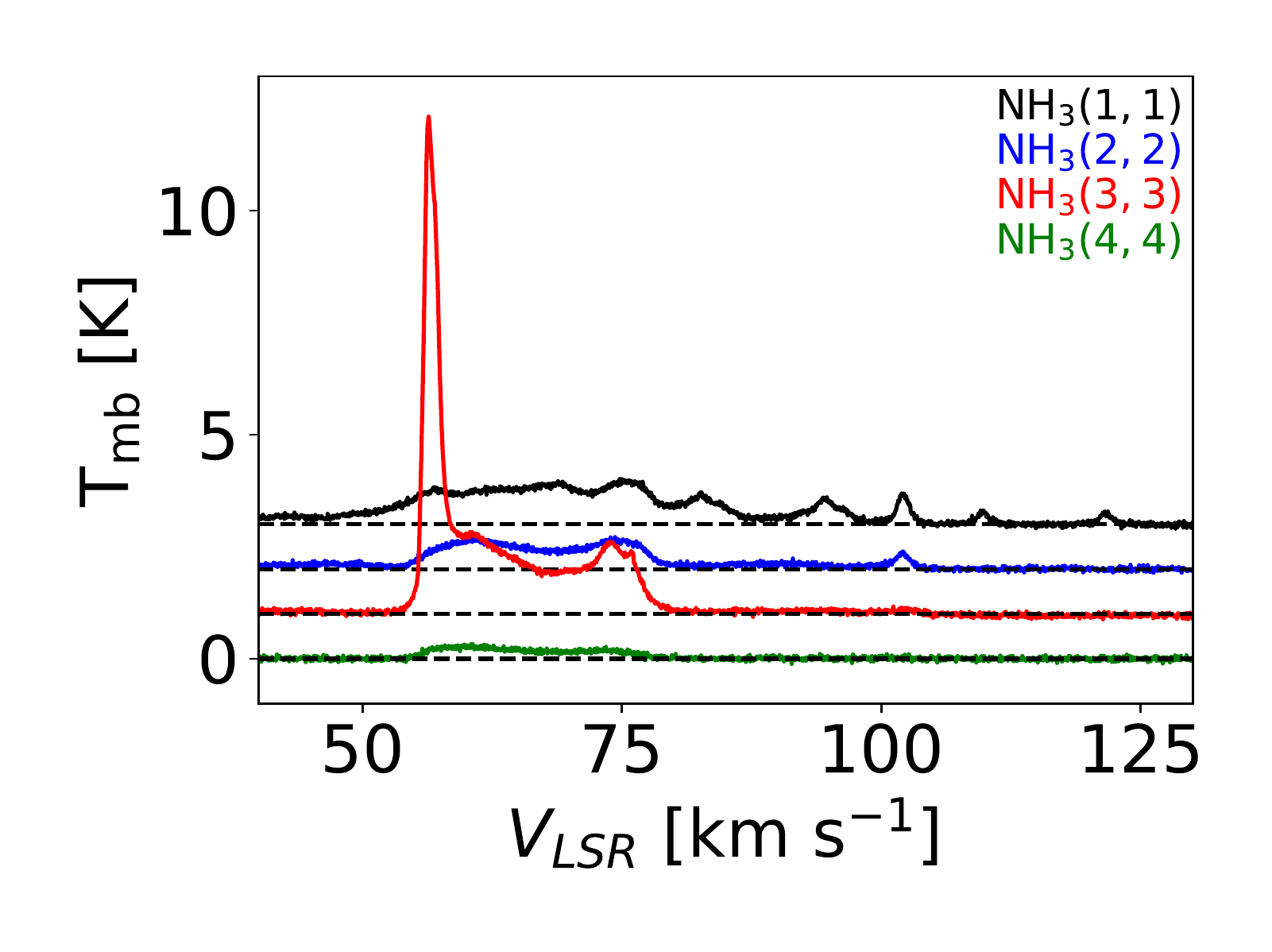}
\caption{RAMPS NH$_3$(1,1) through (4,4) spectra toward the peak of the narrow NH$_3$(3,3) emission. The broad emission from G23.33-0.30 is centered near $V_{\rm{LSR}} = 67$ km s$^{-1}$, while the more typical NH$_3$ emission from a presumably unrelated source peaks at $V_{\rm{LSR}} = 103$ km s$^{-1}$. The emission from G23.33-0.30 exhibits extremely broad line widths, enhanced NH$_3$(3,3)/(1,1) brightness temperature ratio, and an NH$_3$(3,3) maser candidate.}
\label{fig:GBT_specs}
\end{figure}

G23.33-0.30 is a dense molecular filament with extremely broad molecular line widths. Due to the lack of any evidence at infrared wavelengths for embedded star formation, these broad line widths are puzzling. The $\mathrm{H_{2}O}$ Southern Galactic Plane Survey \citep[HOPS; ][]{2011MNRAS.416.1764W} detected extremely broad NH$_3$ line widths and a bright, narrow feature in the NH$_3$(3,3) line profile toward G23.33-0.30. Figure~\ref{fig:GBT_specs} displays Radio Ammonia Mid-Plane Survey \citep[RAMPS; ][]{2018ApJS..237...27H} NH$_{3}$(1,1) through (4,4) spectra toward G23.33-0.30 at higher angular resolution ($\theta \sim 32''$) than that of HOPS ($\theta \sim 120''$). The spectra show the uncommonly broad line emission associated with G23.33-0.30, as well as emission from an unrelated background MC with a local standard of rest (LSR) velocity of $V_{\rm{LSR}} = 103$ km s$^{-1}$ that exhibits more typical NH$_3$ line profiles. G23.33-0.30's molecular line emission has a large velocity dispersion ($\sigma \sim 8 \ \rm{km \ s^{-1}}$), much larger than that of a typical dense MC \citep[$\sigma \sim 0.5-3 \ \rm{km \ s^{-1}}$; ][]{2012ApJ...756...60S}. Moreover, the NH$_3$(3,3) spectrum displays a bright, narrow line component superposed on a fainter, broad component. Unlike the broad component, the bright, narrow line emission in the ortho ($K = 3n$) NH$_3$(3,3) spectrum has no corresponding component in the para ($K \neq 3n$) transitions. Because masers emit at such small spatial scales \citep[e.g., ][]{1992ARA&A..30...75E}, the single-dish observations lack the angular resolution necessary to definitively determine whether the brightness temperature of this narrow feature is large enough to confirm maser emission. Nevertheless, the narrow line width of this component and its appearance solely in the NH$_3$(3,3) spectrum suggests that it is from an NH$_3$(3,3) maser. Compared to other masing transitions, NH$_3$(3,3) masers are exceedingly rare. Indeed, to our knowledge only 15 sources with NH$_3$(3,3) masers have been discovered outside of the Galactic Center (Table~\ref{tab:knownmasers}). Although our understanding of NH$_3$(3,3) maser excitation is incomplete, studies have found that they can be excited by shocks resulting from SNR-MC interactions \citep{2016ApJ...826..189M} or energetic outflows from high-mass protostars \citep{1994ApJ...428L..33M,1995ApJ...439L...9K,1995ApJ...450L..63Z}. While the majority of NH$_3$(3,3) masers have fluxes of $\lesssim1$ Jy, \citet{2011MNRAS.416.1764W} measured a flux of 9.7 Jy for G23.33-0.30's maser candidate, potentially making it the brightest yet detected. The unusually broad line widths may indicate that G23.33-0.30 is experiencing a particularly extreme impact from either a SNR or a high-mass protostellar outflow. 


\begin{deluxetable}{lcr}
\tablecolumns{3}
\tablecaption{Known NH$_3$(3,3) Masers \label{tab:knownmasers}}

\tablehead{\colhead{Source}&
		  \colhead{Flux (Jy)}&
		  \colhead{Reference}}
\startdata
DR21(OH) & 0.260 & \citet{1994ApJ...428L..33M}\\
W51 & 0.230 & \citet{1995ApJ...450L..63Z} \\
NGC 6334 V & 0.114 & \citet{1995ApJ...439L...9K} \\
NGC 6334 I & 0.482 & \citet{1995ApJ...439L...9K} \\
IRAS 20126+4104 & 0.079  & \citet{1999ApJ...527L.117Z} \\
G5.89-0.39 & 0.031 & \citet{2008ApJ...680.1271H} \\
G20.08-0.14N & 0.191 & \citet{2009ApJ...706.1036G} \\
G23.33-0.30 & 9.7 & \citet{2011MNRAS.416.1764W} \\
G30.7206-00.0826 & 5 & \citet{2011MNRAS.418.1689U} \\
G35.03+0.35 & 0.065 & \citet{2011ApJ...739L..16B} \\
G28.34+0.06 & 0.03  & \citet{2012ApJ...745L..30W} \\
W51C & 1.4 & \citet{2016ApJ...826..189M} \\
W44 & 0.07 & \citet{2016ApJ...826..189M} \\
G5.7-0.0 & 0.35 & \citet{2016ApJ...826..189M} \\
G1.4-0.1 & 0.58 & \citet{2016ApJ...826..189M} \\
\enddata
\end{deluxetable}

G23.33-0.30's broad NH$_3$ line emission is associated with a filamentary infrared dark cloud (IRDC) that resides within the giant molecular cloud (GMC) G23.0-0.4, a large ($\sim84\times15$ pc), massive ($\sim5\times10^5 \ \rm{M_{\odot}}$), and dense ($\sim10^3 \ \rm{cm^{-3}}$) filamentary GMC \citep{2015ApJ...811..134S} that hosts multiple generations of high-mass star formation \citep{2014A&A...569A..20M}. In particular, there are several nearby SNRs projected against G23.0-0.4, two of which, G22.7-0.2 \citep{2014ApJ...796..122S} and W41 \citep{2013ApJ...773L..19F,2015ApJ...811..134S}, may be interacting with the GMC. Furthermore, W41 exhibits 20 cm continuum emission, two 1720 MHz OH maser candidates, and extended TeV emission coincident with or adjacent to G23.33-0.30. G23.33-0.30's large peak H$_2$ column density of $N_{H_2} = 1.2\times10^{23}$ cm$^{-2}$ \citep{2016A&A...590A..72P} is similar to that of other high-mass IRDCs, which are thought to be the formation sites of high-mass stars and stellar clusters \citep{2006ApJ...641..389R}. The Co-Ordinated Radio 'N' Infrared Survey for High-mass star formation's \citep[CORNISH; ][]{2012PASP..124..939H} non-detection of an \Hii\ region likely indicates that any high-mass stars forming within G23.33-0.30 are in an embedded pre-stellar or protostellar phase. Considering GMC G23.0-0.4's potential involvement in a SNR-MC interaction, as well as G23.33-0.30's potential capacity for high-mass star formation, both stellar outflows or a SNR-MC interaction remain viable explanations for the excitation of the potential NH$_3$(3,3) maser emission.

To confirm the suspected NH$_3$(3,3) maser emission, determine its excitation conditions, and investigate the nature of the broad NH$_3$ line widths, we have performed followup observations of G23.33-0.30 that probe the filament at the $<0.1$ pc scale. In this paper we present Karl G. Jansky Very Large Array (VLA) observations of the NH$_3$(1,1) through (6,6) inversion lines, Atacama Compact Array (ACA) observations of SiO(5-4), CS(5-4), and 1.3 mm continuum, and Submillimeter Array\footnote{The Submillimeter Array is a joint project between the Smithsonian Astrophysical Observatory and the Academia Sinica Institute of Astronomy and Astrophysics, and is funded by the Smithsonian Institution and the Academia Sinica.} (SMA) observations of $^{13}$CO(2-1), C$^{18}$O(2-1), and 1.3 mm continuum. In Section~\ref{sec:obs} we describe these observations and the reduction of the data, in Section~\ref{sec:res} we present the results, in Section~\ref{sec:ana} we analyze the data, in Section~\ref{sec:disc} we discuss the analysis, and in Section~\ref{sec:con} we provide our conclusions. 

\section{Observations and Data Reduction}
\label{sec:obs}

We have observed G23.33-0.30 using the VLA, operated by the National Radio Astronomy Observatory\footnote{The National Radio Astronomy Observatory (NRAO) is a facility of the National Science Foundation operated under cooperative agreement by Associated Universities, Inc.}, the ACA, and the SMA. Table~\ref{tab:newdata} provides a summary of the continuum and spectral line data analyzed in this work and Sections~\ref{subsec:vla}, \ref{subsec:aca}, and \ref{subsec:sma} describe the calibration and reduction of these data. We also display archival data from several surveys, which are summarized in Table~\ref{tab:archdata}.

\begin{deluxetable}{ccccccccc}

\tabletypesize{\tiny}
\tablecolumns{9}
\tablecaption{New Observations \label{tab:newdata}}
\tablehead{\colhead{Telescope}&
		 \colhead{Date}&
		 \colhead{Transition}&
		 \colhead{$\nu_{0}$}&
		 \colhead{$\theta_{P}$}&
		 \colhead{$\theta_{maj} \times \theta_{min}$}&
		 \colhead{$\Delta V_{BW}$}&
		 \colhead{$\Delta V_{chan}$}&
		 \colhead{$\sigma_{noise}$} \\
		 \colhead{}&
		 \colhead{}&
		 \colhead{}&
		 \colhead{(GHz)}&
		 \colhead{(arcmin)}&
		 \colhead{(arcsec)}&
		 \colhead{(km s$^{-1}$)}&
		 \colhead{(km s$^{-1}$)}&
		 \colhead{(mJy beam$^{-1}$)}}
\startdata
SMA & 2016 Jun 20 & $^{13}$CO(2-1) & 220.39868 & 0.8 & $4.1 \times 2.2$ & 345 & 0.17 & 82.5\\ 
$-$ & $-$ & C$^{18}$O(2-1) & 219.56035 & 0.8 & $4.1 \times 2.2$ & 173 & 0.17 & 92.0\\ 
$-$ & $-$ & 1.3 mm cont. & \nodata & 0.8 & $4.1 \times 2.2$ & 10435 &  \nodata & 1.1\\ 
VLA-A & 2016 Oct 7 & NH$_3$(3,3) & 23.87013 & 1.9 & $0.13 \times 0.09$  & 50 & 0.39 & 8.1\\
VLA-D & 2017 Apr 14 & NH$_3$(1,1) & 23.69450 & 1.9 & $3.8 \times 2.7$ & 51 & 0.40 & 1.6 \\
$-$ & $-$ & NH$_3$(2,2) & 23.72263 & 1.9 & $3.7 \times 2.7$  & 51 & 0.39 & 1.3\\
$-$ & $-$ & NH$_3$(3,3) & 23.87013 & 1.9 & $3.7 \times 2.7$  & 50 & 0.39 & 0.6\\
$-$ & $-$ & NH$_3$(4,4) & 24.13942 & 1.9 & $3.4 \times 2.6$  & 50 & 0.39 & 1.4\\
$-$ & $-$ & NH$_3$(5,5) & 24.53299 & 1.8 & $3.4 \times 2.6$ & 49 & 0.38 & 1.1\\
$-$ & $-$ & NH$_3$(6,6) & 25.05603 & 1.8 & $3.3 \times 2.6$  & 48 & 0.37 & 1.6\\
ACA & 2017 Jul 10-13 & SiO(5-4) & 217.10498 & 0.8 & $7.2 \times 4.5$  & 689 & 0.34 & 20.2\\  
$-$ & $-$ & 1.3 mm cont. & \nodata & 0.8 & $6.6 \times 4.0$  & 5451 & \nodata & 3.7\\ 
$-$ & 2017 Jul 8 & CS(5-4) & 244.93556 & 0.7 & $6.0 \times 4.2$  & 261 & 0.29 & 22.9\\ 
$-$ & $-$ & 1.3 mm cont. & \nodata & 0.7 & $6.4 \times 4.1$  & 4984 & \nodata & 4.3\\ 
\enddata
\tablecomments{$\nu_{0}$ is the rest frequency of the spectral line, $\theta_{P}$ is the full width at half maximum (FWHM) size of the primary beam, $\theta_{maj} \times \theta_{min}$ is the FWHM size of the synthesized beam, $\Delta V_{BW}$ is the spectral bandwidth, and $\Delta V_{chan}$ is the spectral resolution.}
\end{deluxetable}

\begin{deluxetable}{cccccccc}

\tabletypesize{\footnotesize}
\tablewidth{0pt}
\tablecolumns{8}
\tablecaption{Archival Data \label{tab:archdata}}
\tablehead{\colhead{Survey}&
		 \colhead{Telescope}&
		 \colhead{Wavelength/Energy}&
		 \colhead{Spatial Resolution}&
		 \colhead{Reference} \\
		 \colhead{}&
		 \colhead{}&
		 \colhead{}&
		 \colhead{}&
		 \colhead{}}
\startdata
GLIMPSE & \textit{Spitzer} & 3.6 $\mu$m &  $<2''$ &  1, 2 \\
GLIMPSE & \textit{Spitzer} & 8 $\mu$m &  $<2''$ &  1, 2  \\
MIPSGAL &  \textit{Spitzer} & 24 $\mu$m &  6$''$ &  3 \\
MAGPIS & VLA & 20 cm & $5\farcs4 \times 6\farcs2$ &  4 \\ 
VGPS & VLA & 21 cm & $1' \times 1'$&  5 \\
GRS & FCRAO 14 m & $^{13}$CO(1-0) - 2.7 mm & 46$''$ &  6 \\
H.E.S.S. Survey of Inner Galaxy & H.E.S.S. & $0.1 - 100$ TeV & $\sim 0\fdg1$ & 7 \\
\enddata
\tablecomments{References: (1) \citealp{2003PASP..115..953B}; (2) \citealp{2009PASP..121..213C}; (3) \citealp{2009PASP..121...76C}; (4) \citealp{2006AJ....131.2525H}; (5) \citealp{2006AJ....132.1158S}; (6) \citealp{2006ApJS..163..145J}; (7) \citealp{2006ApJ...636..777A}.}
\end{deluxetable}

\subsection{VLA Observations}
\label{subsec:vla}
We observed G23.33-0.30 using the VLA in the D array configuration for a seven hour track. We performed the bandpass and flux calibration of the data using observations of J1331+305 (3C286) and we performed the phase calibration using periodic observations of J1851+0035. The calibration and imaging of the data were performed using CASA\footnote{https://casa.nrao.edu/} 5.1.1-5 and we imaged the data using CASA's \texttt{clean} algorithm with Briggs weighting and the robustness parameter set to 0.5.  

G23.33-0.30 was also recently observed by \citet{KS} using the VLA in the A array configuration, which only detected NH$_3$(3,3) emission. The A array observation used the same calibrators as the D array, but the bandpass/flux calibrator was partially resolved by the A array's long baselines. Consequently, the fluxes measured from the A array data are lower limits. The spectral band for the A array observations was shifted to lower $V_{LSR}$ compared to the D array, so the D array observations were able to detect emission up to $V_{LSR} = 90$ km s$^{-1}$, while the A array observations could only detect emission up to $V_{LSR} = 75$ km s$^{-1}$. The A array data were also reduced using CASA 5.1.1-5 and were imaged using CASA's \texttt{clean} algorithm with natural weighting.  

\subsection{ACA Observations}
\label{subsec:aca}

We observed G23.33-0.30 with the ACA using three pointings and five 50 min execution blocks in two spectral setups. The first spectral setup was used to observe SiO(5-4) and executed four times and the second spectral setup was used to observe CS(5-4) and was executed once. For both spectral setups we performed the bandpass calibration using observations of J1924-2914 and the phase calibration using periodic observations of J1743-0350. For the first spectral setup we performed the flux calibration using observations of J1733-1304 and for the second spectral setup we used observations of J1751+0939. The ACA data were calibrated by the ALMA data reduction pipeline and imaged using CASA 4.7.2. We carried out imaging using CASA's \texttt{tclean} algorithm with Briggs weighting and the robustness parameter set to 0.5.

\subsection{SMA Observations}
\label{subsec:sma}
We observed G23.33-0.30 using the SMA for an eight hour track in the compact configuration. We performed the bandpass calibration using observations of 3C454.3, the flux calibration using observations of Neptune, and the phase calibration using periodic observations of 1743-038. We calibrated the data using MIR\footnote{https://www.cfa.harvard.edu/rtdc/SMAdata/process/mir/}, an IDL-based data reduction software package, and converted the calibrated data to MIRIAD\footnote{https://www.atnf.csiro.au/computing/software/miriad/} format for imaging using the \texttt{mir2miriad} procedure. We imaged the data with MIRIAD 4.3.8 using MIRIAD's \texttt{clean} algorithm with Briggs weighting and the robustness parameter set to 1.

\section{Results}
\label{sec:res}

\begin{figure}[h]
\centering
\includegraphics[scale=0.18]{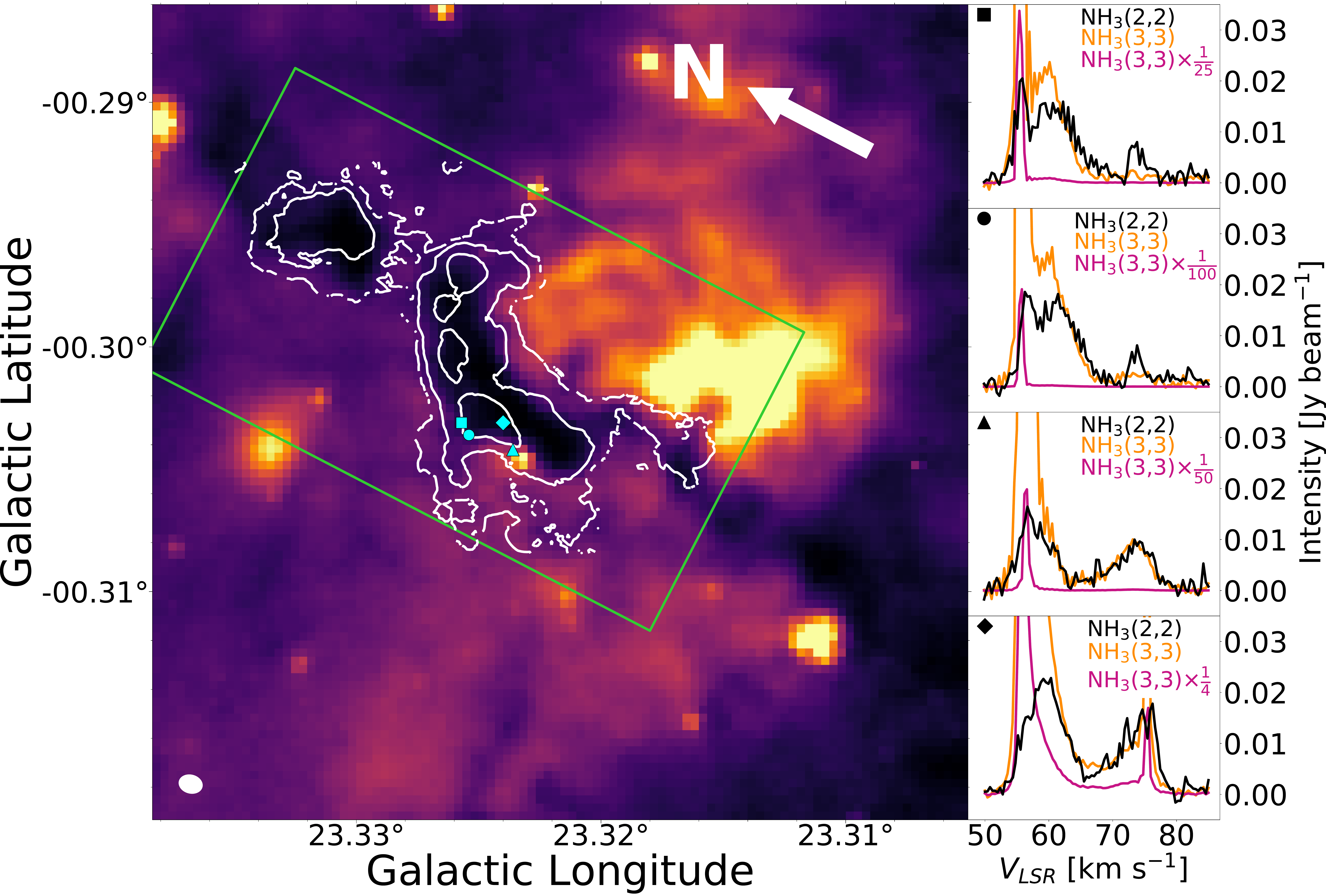}
\caption{Left: GLIMPSE 8 $\mu$m map of G23.33-0.30 with VLA $\mathrm{NH_3}$(2,2) integrated intensity contours overlaid at 10, 50, and 150 mJy beam$^{-1}$ km s$^{-1}$. The VLA D Array beam is shown in the lower left corner of the map and the arrow in the upper right corner points to the north celestial pole. The green rectangle shows the region used to make the position-velocity diagram in Figure~\ref{fig:dec-vlsr}. The symbol markers indicate the positions of the NH$_3$(3,3) maser candidates. The thermal NH$_3$ emission traces an IRDC filament. Right: VLA D array NH$_3$(2,2) and (3,3) spectra toward the four maser candidates, where the symbol markers in the upper left of each plot correspond to those in the GLIMPSE 8 $\mu$m map. The NH$_3$(3,3) spectra are presented both at their true amplitude and scaled for comparison.}
\label{fig:MIR_VLA_masers}
\end{figure}

Figure~\ref{fig:MIR_VLA_masers} shows the 8 $\mu$m Galactic Legacy Infrared Midplane Survey Extraordinaire  \citep[GLIMPSE; ][]{2003PASP..115..953B,2009PASP..121..213C} image of G23.33-0.30 with the VLA NH$_3$(2,2) integrated intensity overlaid as contours. Clearly, the thermal NH$_3$(2,2) emission traces the IRDC filament, while a gap in the 8 $\mu$m extinction ($l = 23.328^{\circ}$) corresponds to a gap in the NH$_3$(2,2) emission. The $\sim 3''$ spatial resolution of the D array has resolved the bright, narrow NH$_3$(3,3) line at $V_{LSR}\sim56$ km s$^{-1}$ detected by HOPS and RAMPS into two point-like sources, while also revealing another potential maser at $V_{LSR}\sim76$ km s$^{-1}$, $<10''$ away from the other two sources. The VLA A array observations of G23.33-0.30 \citep{KS} resolved one of the maser candidates detected by the D array into two sources. The positions of the four maser candidates are shown with symbol markers in Figure~\ref{fig:MIR_VLA_masers}, which also displays the NH$_3$(2,2) and NH$_3$(3,3) spectra toward each source. The spectra show that the suspected NH$_3$(3,3) maser emission is much narrower and brighter than the thermal NH$_3$(2,2) and (3,3) emission. Three of the maser candidates have velocities near the peak of the narrowest NH$_3$(2,2) component at 56 km s$^{-1}$, while the faintest is found near a component peaking at 77 km s$^{-1}$.

\begin{figure}
\centering
\includegraphics[scale=0.3]{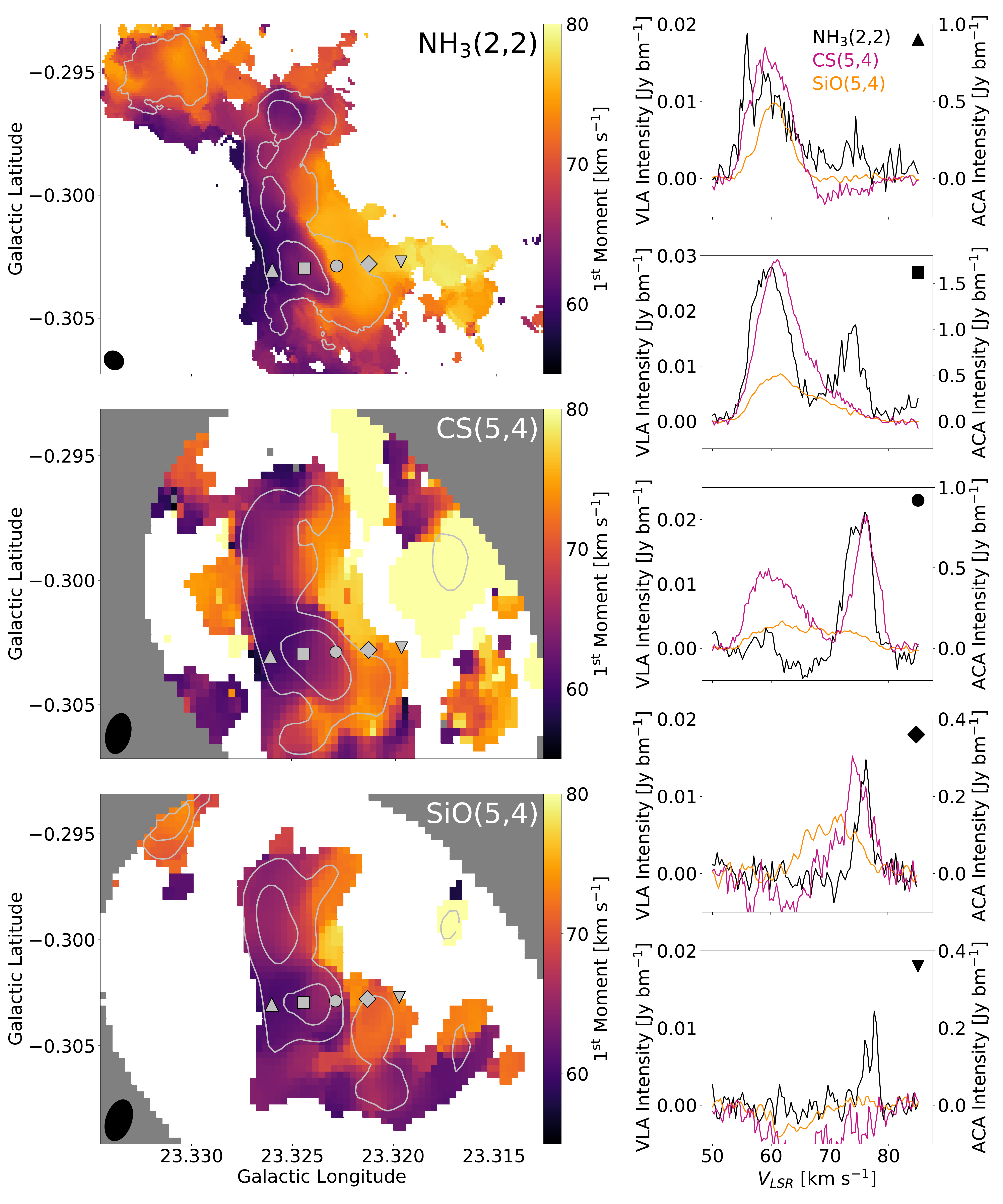}
\caption{Left: NH$_3$(2,2) (top), CS(5-4) (middle), and SiO(5-4) (bottom) $1^{\rm{st}}$ moment maps overlaid with integrated intensity (moment 0) contours at 0.05 and 0.15, 2 and 8, and 1 and 4 Jy beam$^{-1}$ km s$^{-1}$, respectively. The symbols overlaid correspond to the locations of the spectra to the right. Right: NH$_3$(2,2) (black), CS(5-4) (magenta), and SiO(5-4) (orange) spectra from the positions indicated by the symbol markers in the left panels. The left axis corresponds to the NH$_3$(2,2) spectra, while the right axis corresponds to the CS(5-4) and SiO(5-4) spectra. These data reveal a velocity discontinuity across the width of the filament.}
\label{fig:NH3_SiO_CS_kin}
\end{figure}

Figure~\ref{fig:NH3_SiO_CS_kin} illustrates the unusual kinematics in G23.33-0.30. The left panels show the NH$_3$(2,2), CS(5-4), and SiO(5-4) 1$^{\rm{st}}$ moment maps of the filament, with their respective integrated intensity contours overlaid, while the right panels show spectra taken across the width of the filament. The NH$_3$(2,2) and CS(5-4) data reveal a velocity discontinuity between a broad line component peaking at $V_{\rm{LSR}} \sim 60$ km s$^{-1}$ that is associated with the left edge of the filament and a narrower component at $V_{\rm{LSR}} \sim 77$ km s$^{-1}$ that is associated with the right side. The NH$_3$(2,2) data exhibit an additional narrow velocity component at $V_{\rm{LSR}} \sim 56$ km s$^{-1}$, peaking near the velocities of three of the maser candidates. The SiO(5-4) emission is extremely broad and peaks primarily at $V_{\rm{LSR}} < 76$ km s$^{-1}$. With the ACA's large spectral bandwidth, we are also able to detect emission near $V_{\rm{LSR}}=100$ km s$^{-1}$. We do not show this emission in any figures, since it is associated with two background sources unrelated to G23.33-0.30. 

\begin{figure}[h]
\centering
\includegraphics[scale=0.23]{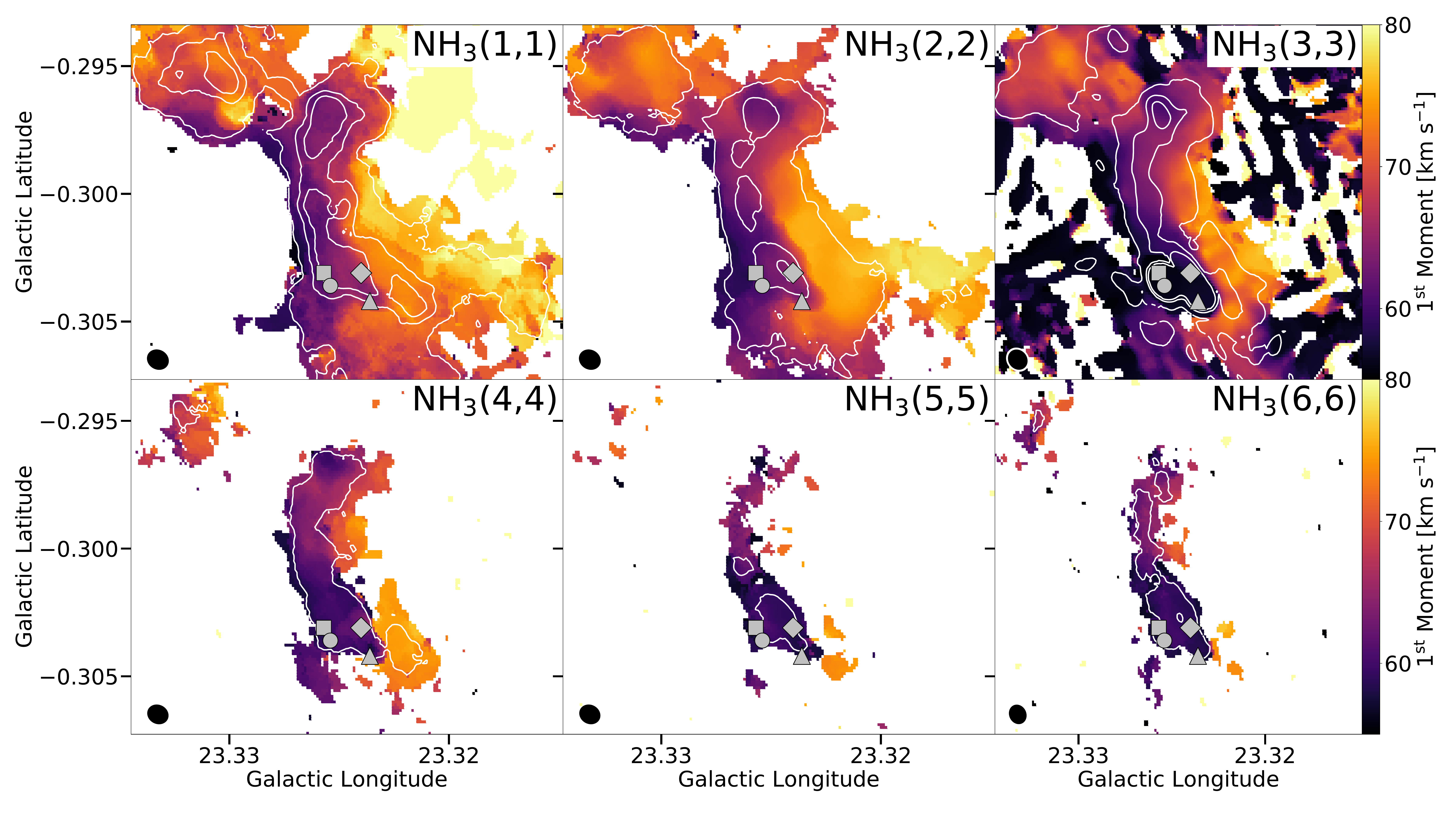}
\caption{NH$_3$(1,1) (top left), (2,2) (top middle), (3,3) (top right), (4,4) (bottom left), (5,5) (bottom middle), and (6,6) (bottom right) $1^{\rm{st}}$ moment maps overlaid with integrated intensity contours at 25, 150, and 250 mJy beam$^{-1}$ km s$^{-1}$. The silver symbol markers indicate the positions of the NH$_3$(3,3) masers and the synthesized beam for each transition is shown in the lower left of the map.}
\label{fig:G23_11-66}
\end{figure}

G23.33-0.30 exhibits bright emission from all of the observed NH$_3$ inversion transitions. Figure~\ref{fig:G23_11-66} displays the NH$_3$(1,1) through (6,6) $1^{\rm{st}}$ moment maps with integrated intensity contours overlaid. The linear features in the NH$_3$(3,3) map are a result of cleaning artifacts, not real emission. The emission from the higher energy transitions (NH$_3$(4,4) through (6,6)) is strongest at lower $V_{\rm{LSR}}$ and displays similarly broad line widths as the 60 km s$^{-1}$ NH$_3$(2,2) velocity component. The highest energy transition, NH$_3$(6,6), features particularly bright emission compared to NH$_3$(5,5), and even displays amplitude ratios of NH$_3$(6,6)/(4,4)$>$1 near the peak of the NH$_3$(6,6) emission. In contrast to the NH$_3$(1,1) and (2,2) data, the NH$_3$(4,4) through (6,6) emission peaks only at $V_{\rm{LSR}}\leq76$ km s$^{-1}$, but exhibits a velocity discontinuity between components peaking at $V_{\rm{LSR}} \sim 59$ and 75 km s$^{-1}$.

\begin{figure}[h]
\centering
\includegraphics[scale=0.35]{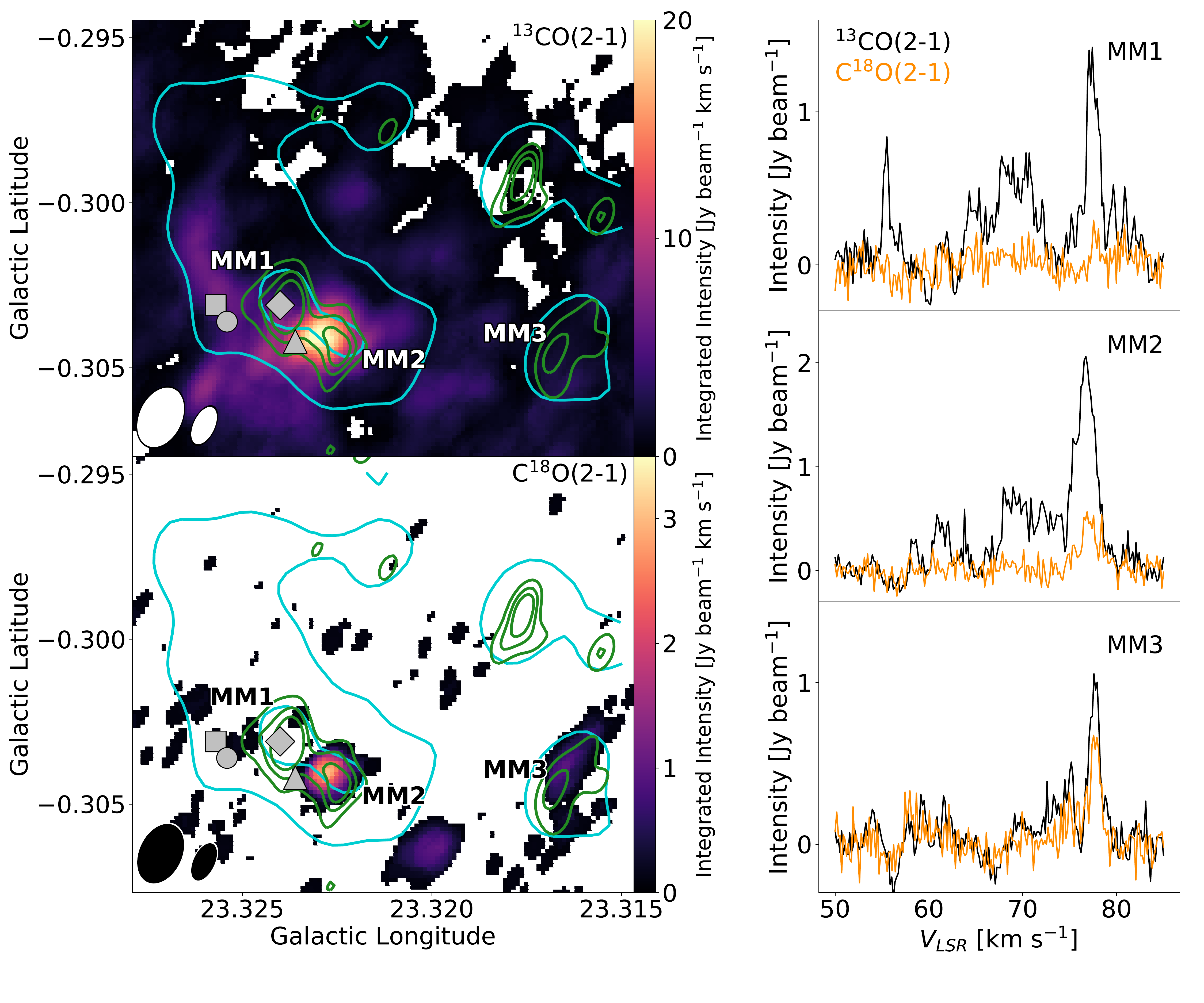}
\caption{Left: Color shows the SMA $^{13}$CO (top) and C$^{18}$O (bottom) integrated intensity maps. Overlaid are 1.3 mm continuum contours from the ACA (cyan) at 4 and 30 mJy beam$^{-1}$ and the SMA (green) at 3.5, 7, and 10.5 mJy beam$^{-1}$. The lowest contours correspond to a 5$\sigma$ detection in both cases. The silver symbol markers indicate the positions of the masers. The continuum data reveal three compact cores associated with the filament: MM1, MM2, and MM3. Right: SMA $^{13}$CO (black) and C$^{18}$O (orange) spectra toward MM1, MM2, and MM3. The brightest line in each spectrum peaks at $V_{\rm{LSR}} \sim 77$  km s$^{-1}$.}
\label{fig:SMA_CO}
\end{figure}

Figure~\ref{fig:SMA_CO} shows the SMA $^{13}$CO(2-1), C$^{18}$O(2-1), 1.3 mm continuum, and ACA 1.3 mm continuum data.  Due to the ACA's good $uv$ coverage at shorter baselines, the continuum emission detected by the ACA traces the larger scale filament, while the SMA continuum observations are primarily sensitive to the compact continuum cores. Three of the continuum cores, MM1, MM2, and MM3, lie along the filament. 
Because the bright mm core west of the filament is coincident with emission near $V_{\rm{LSR}}=100$ km s$^{-1}$, it is not associated with G23.33-0.30. The SMA $^{13}$CO(2-1) and C$^{18}$O(2-1) emission peaks at $V_{\rm{LSR}} \sim 77$  km s$^{-1}$ between MM1 and MM2. The $^{13}$CO(2-1) and C$^{18}$O(2-1) spectra peak at $V_{\rm{LSR}} \sim 77$  km s$^{-1}$ toward each of the cores associated with G23.33-0.30. There is also $^{13}$CO(2-1) emission at lower $V_{\rm{LSR}}$ appearing mainly on the eastern edge of the filament in the range $V_{\rm{LSR}} = 56-70$  km s$^{-1}$. 

\begin{figure}[h]
\centering
\includegraphics[scale=0.5]{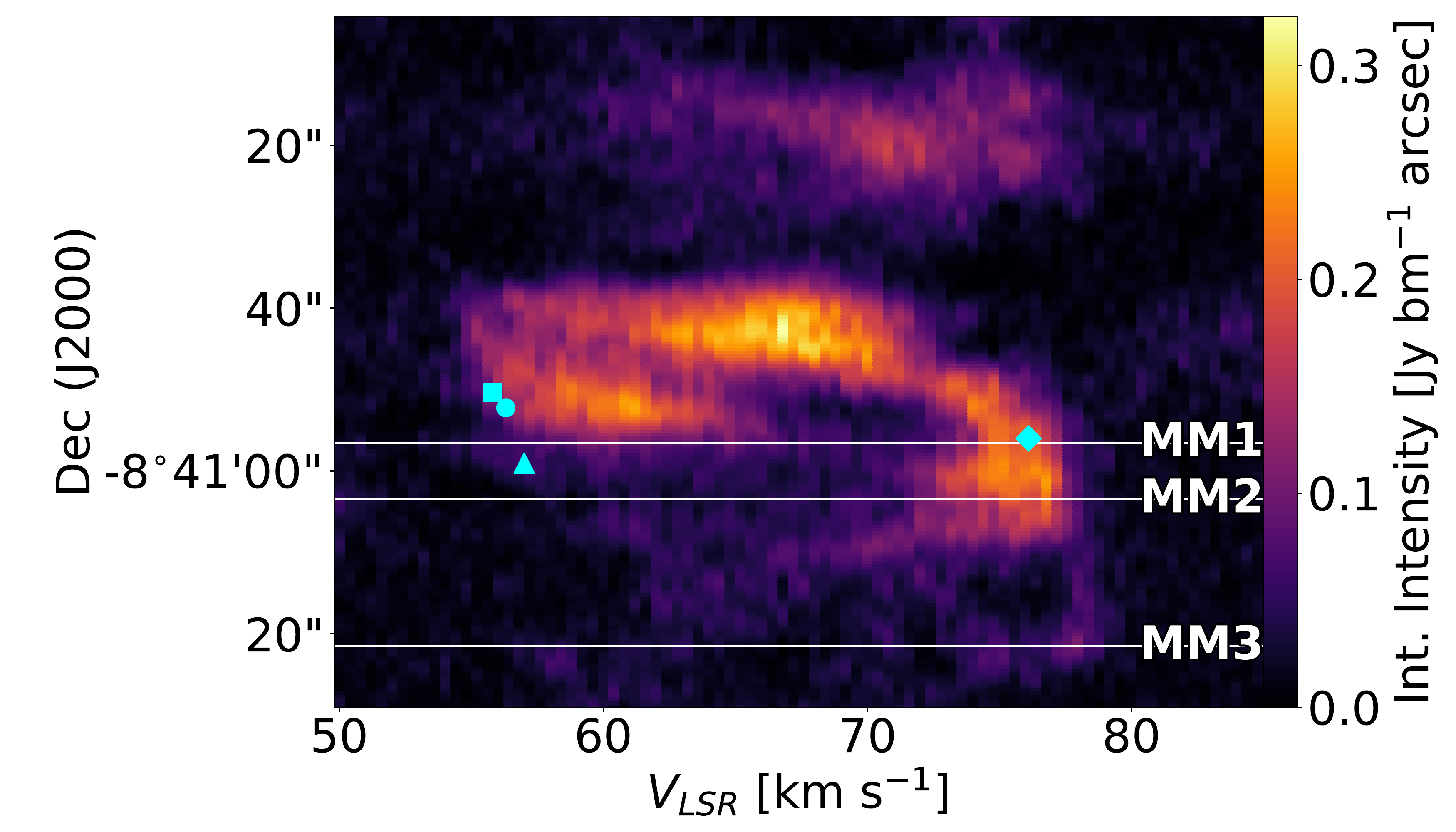}
\caption{Plot of Declination vs. $V_{\rm{LSR}}$ made from the NH$_3$(2,2) intensity integrated over the Right Ascension axis. The region over which we performed the integration is shown with a green box in Figure~\ref{fig:MIR_VLA_masers}. Because G23.33-0.30 is oriented roughly north to south, this displays the gas kinematics along the filament's length. The cyan symbol markers indicate the positions and velocities of the NH$_3$(3,3) masers and the white horizontal lines mark the positions of the molecular cores MM1, MM2, and MM3.}
\label{fig:dec-vlsr}
\end{figure}

Figure~\ref{fig:dec-vlsr} displays the NH$_3$(2,2) intensity integrated along lines of constant Right Ascension. Because the filament points roughly north to south, the figure shows the variation of the gas kinematics along G23.33-0.30's length. We have also overplotted the positions and velocities of the NH$_3$(3,3) maser candidates and the positions of the continuum sources MM1, MM2, and MM3. While the vast majority of the emission is at $V_{\rm{LSR}} < 78$ km s$^{-1}$ and displays broad line widths, a small portion of the filament between MM2 and MM3 exhibits narrow line emission peaking at $V_{\rm{LSR}} \sim 78$ km s$^{-1}$. This narrow line emission corresponds to the southernmost section of the filament detected in NH$_3$(2,2), but not SiO(5-4). The gas north of this narrow line emission is at lower $V_{\rm{LSR}}$ and composed of two components: a narrow component at $V_{\rm{LSR}} \sim 56$ km s$^{-1}$ that is spatially and spectrally coincident with three of the NH$_3$(3,3) maser candidates (Fig.~\ref{fig:MIR_VLA_masers}), and a turbulent component peaking at $V_{\rm{LSR}} = 60-76$ km s$^{-1}$. The section of the filament associated with the NH$_3$(3,3) maser candidate at $V_{\rm{LSR}}=76.4$ km s$^{-1}$ is less turbulent and less blueshifted than the northern part of the filament. Figure~\ref{fig:dec-vlsr} also reveals slight deficits in emission in the range $67 < V_{\rm{LSR}} < 75$ km s$^{-1}$ at the positions of MM1 and MM2, while the NH$_3$ emission toward MM3 reveals a velocity discontinuity between components peaking at $V_{\rm{LSR}} =74$ km s$^{-1}$ and $V_{\rm{LSR}} =78$ km s$^{-1}$.

\section{Analysis}
\label{sec:ana}

\subsection{NH$_3$(3,3) Maser Emission}
We determined the position and flux of each NH$_3$(3,3) maser candidate by first fitting spectra to determine line amplitudes and then fitting these amplitude maps to estimate the positions. We used the Python Markov Chain Monte Carlo (MCMC) fitting package \texttt{emcee} \citep{2013PASP..125..306F} to fit both the narrow line emission in the data cube and the resulting maps of the line amplitudes. We fit the narrow line emission with a Gaussian model and fit the amplitude maps with a model of the synthesized beam, since we expect maser emission to be unresolved. Table~\ref{tab:masers} presents the A array fit results for the sources peaking at $56-57$ km s$^{-1}$ and the D array fit results for the source peaking at 76 km s$^{-1}$, as well as the symbol markers corresponding to each source shown in Figures~\ref{fig:MIR_VLA_masers}, \ref{fig:G23_11-66}, \ref{fig:SMA_CO}, and \ref{fig:dec-vlsr}. We display the best-fit values for each source's Galactic coordinate position ($l,b$), flux ($I_{\nu}$), LSR velocity ($V_{\rm{LSR}}$), velocity dispersion ($\sigma$), and line brightness temperature ($\Delta T_B$).

\begin{deluxetable}{lcccc}

\tabletypesize{\footnotesize}
\tablecolumns{5}
\tablecaption{VLA NH$_3$(3,3) Maser Properties \label{tab:masers}}

\tablehead{\colhead{}&
		  \colhead{Maser 1}&
		  \colhead{Maser 2}&
		  \colhead{Maser 3}&
		  \colhead{Maser 4}}
\startdata
Symbol & $\Circle$ & $\triangle$ & $\Box$ & $\Diamond$ \\
Array & A & A & A & D \\
$l$ (deg) & 23.325713$\pm$0.000001 & 23.325393$\pm$0.000001 & 23.323564$\pm$0.000001 & 23.323966$\pm$0.000002 \\
$b$ (deg) &  -0.303063$\pm0.000001$ & -0.303566$\pm$0.000001 & -0.304137$\pm$0.000001  & -0.303085$\pm$0.000002 \\
$I_{\nu}$ (Jy) & $1.312\pm0.086$ & $0.562\pm0.049$ & $0.453\pm0.033$ & $0.078 \pm 0.001$ \\
$V_{\rm{LSR}}$ (km s$^{-1}$)  & $56.319\pm0.002$ & $57.025\pm0.007$ & $55.754\pm0.008$  & $76.375\pm 0.003$ \\
$\sigma$ (km s$^{-1}$) & $0.254\pm0.003$ & $0.242\pm0.007$ & $0.213\pm0.005$ & $0.187 \pm 0.004$\\
$\Delta T_B$ (K) & $171300\pm11230$ & $73380\pm6398$ & $59150\pm4309$ & $16.7\pm0.2$ \\
\enddata
\end{deluxetable}

Maser emission occurs in gas with a population inversion, with more molecules in the upper state of a transition than  expected in local thermodynamic equilibrium (LTE). Population inversions in the NH$_3$(3,3) transition are a result of collisions with H$_2$ molecules \citep{1983A&A...122..164W}. The brightness temperature of a spectral line is given by ${\Delta T_B = (T_{ex} - T_{bkg})(1 - e^{-\tau_{\nu}})}$, where $T_{ex}$ is the transition's excitation temperature, $T_{bkg}$ is the background temperature, and $\tau_{\nu}$ is the optical depth at frequency $\nu$. If the molecular transition is in LTE with gas at temperature $T_{gas}$ and the background radiation is dominated by the Cosmic Microwave Background (CMB), then ${\Delta T_B \leq (T_{gas} - T_{CMB})}$. On the other hand, a nonthermal population inversion produces negative values of $\tau_{\nu}$ and $T_{ex}$, resulting in $\Delta T_B \gg T_{gas}$. Given the low temperatures expected in G23.33-0.30 (${T_{gas} \ll 10^4 \ \rm{K}}$), the three masers detected by the A array all exhibit ${\Delta T_B \gg T_{gas}}$ , confirming their nonthermal nature. Because the source at $V_{\rm{LSR}}=76$ km s$^{-1}$ was outside of the velocity range of the A array data, the D array data provide our only measurement of its brightness temperature. This source is much fainter than the NH$_3$(3,3) masers at $V_{\rm{LSR}} \sim 56-57$ km s$^{-1}$, exhibiting a D array brightness temperature of only 17 K, which is comparable to the temperatures typically measured in molecular clouds. Although the D array brightness temperature cannot prove this source's nonthermal origin, it is approximately twice as bright as the peak thermal emission and exhibits a narrow line width, similar to those of the confirmed masers. Moreover, the NH$_3$(3,3) maser candidate has no corresponding velocity feature in any of the other NH$_3$ spectra. Consequently, we assume that the emission is nonthermal and will refer to this source as an NH$_3$(3,3) maser. Thus, G23.33-0.30 hosts four NH$_3$(3,3) masers: three associated with gas corresponding to the narrow NH$_3$(2,2) velocity component near 57 km s$^{-1}$, and a fourth associated with the asymmetric line emission peaking at 77 km s$^{-1}$.

\subsection{Thermal NH$_3$ Emission}

\begin{figure}[h]
\centering
\includegraphics[scale=.23]{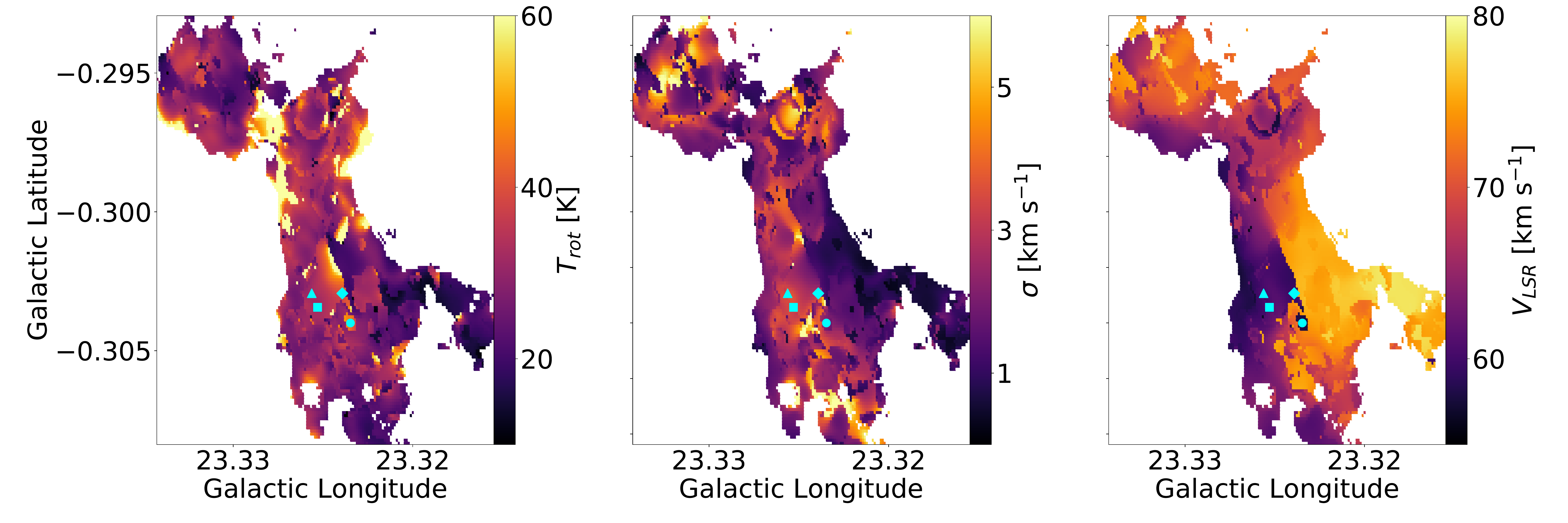}
\caption{Maps of the NH$_3$ model best-fit values of $T_{\rm{rot}}$, $\sigma$, and $V_{\rm{LSR}}$ with the positions of the NH$_3$(3,3) masers overlaid. The gas at the pre-shock velocity is cold and has a low velocity dispersion, while the turbulent component is hot and has $\sigma > 1$ km s$^{-1}$.}
\label{fig:NH3_par_maps}
\end{figure}

In addition to accelerating gas, shocks can heat and add turbulence to the entrained gas component. Using the NH$_3$ modeling methods described by \citet{2018ApJS..237...27H}, we employed a PySpecKit \citep{2011ascl.soft09001G} LTE NH$_3$ model to investigate whether the gas properties in G23.33-0.30 indicate a shock. We first derived NH$_3$ rotational temperatures ($T_{\rm{rot}}$), velocity dispersions ($\sigma$), and LSR velocities ($V_{\rm{LSR}}$) from the NH$_3$(1,1) and (2,2) data cubes. In order to exclude emission that would be too faint to provide accurate derived quantities, we only fit pixels that had NH$_3$(2,2) integrated intensities greater than 10 mJy beam$^{-1}$ km s$^{-1}$. Figure~\ref{fig:NH3_par_maps} displays maps of the best-fit parameter values for the velocity component with the larger NH$_3$(1,1) through (2,2) integrated intensity. The maps are overlaid with the positions of the NH$_3$(3,3) masers, which we expect to reside at the locations of shock fronts. Although the presence of broad, overlapping, and asymmetric line shapes did not allow for accurate fit results over the full map, it is clear that the gas at higher $V_{\rm{LSR}}$ is generally colder and has a lower velocity dispersion than that at lower $V_{\rm{LSR}}$. The southernmost section of the filament, which has the largest measured $V_{\rm{LSR}}$, was detected in NH$_3$(2,2) but not SiO(5-4). It also corresponds to the region that is coldest and has the lowest velocity dispersion, likely indicating that the gas in this portion of the filament is unshocked and that $V_{\rm{LSR}} \sim 78$ km s$^{-1}$ is the filament's pre-shock LSR velocity. While our analysis of the NH$_3$(1,1) and (2,2) data implied only moderate heating of the turbulent component ($\Delta T \sim 30-40$ K), the relatively bright emission from the higher energy transitions NH$_3$(4-4) through (6,6) indicates the presence of a hotter component to which the NH$_3$(1,1) and (2,2) amplitudes are insensitive. Consequently, we also performed fits using all of the observed para-NH$_3$ lines: NH$_3$(1,1), (2,2), (4,4), and (5,5). However, the best-fit models of the para-NH$_3$ lines often featured NH$_3$(1,1) amplitudes that were smaller than those observed in the data and NH$_3$(2,2) amplitudes larger than observed in the data. The fact that a single-temperature NH$_3$ model could not reproduce the para-NH$_3$ amplitudes suggests that there exists at least two temperature components. Thus, we performed fits using only NH$_3$(4,4) and (5,5) to better determine the temperature of the hotter component. The rotational temperatures based only on the NH$_3$(4,4) and (5,5) emission are $T_{rot}(4,4; 5,5)\sim 40-200$ K, much higher than those derived from NH$_3$(1,1) and (2,2). Thus, the shock has deposited significant thermal ($T_{pre-shock} \sim 10-20$ K vs. $T_{post-shock} \sim 40-200$ K) as well as turbulent ($\sigma_{pre-shock} < 1$ km s$^{-1}$ vs. $\sigma_{post-shock} \sim 1-5$ km s$^{-1}$) energy into the filament. 

\subsection{Virial Analysis}
\label{subsec:vir}

We detected the 1.3 mm dust continuum emission from three compact sources associated with the filament, MM1, MM2, and MM3 (Fig.~\ref{fig:SMA_CO}), which represent molecular cores embedded within G23.33-0.30. We used an MCMC routine to fit the positions and sizes of the cores. The compact continuum emission from these cores is superposed on more extended emission. Consequently, we modeled the SMA continuum emission as the superposition of two elliptical Gaussians convolved with the SMA synthesized beam and estimated their positions ($l,b$) and sizes ($R_1 \times R_2$). This analysis implies that MM1 and MM2 are at most barely resolved by the SMA synthesized beam, so their best-fit sizes are upper limits. Table~\ref{tab:cont} displays the fit results. 

The collapse of a molecular core depends on its turbulent energy content. Due to the added support against gravity, gas that is highly turbulent will have more difficulty collapsing to form stars than will gas with lower levels of turbulence. Thus, the line of sight velocity dispersion, along with the core masses, can inform whether the detected cores can collapse to form stars. In order to evaluate the fate of G23.33-0.30 and its associated molecule cores, we performed a virial analysis using the SMA continuum data and $^{13}$CO(2-1) spectra. Neglecting magnetic fields and external pressure, the gravitational stability of a molecular core is dictated by the core mass ($M_{core}$) and the virial mass ($M_{vir}$) in the form of the virial parameter $\alpha = \frac{M_{vir}}{M_{core}}$, where $\alpha<1$ indicates the potential for collapse. We determined $M_{core}$ using the following equation from \citet{1983QJRAS..24..267H}
\begin{equation}
M_{core} = \mathbb{R} \frac{F_{\nu} D^2}{\kappa_{\nu} B_{\nu}(T)}, 
\end{equation}
where $\mathbb{R}$ is the gas-to-dust mass ratio, $F_{\nu}$ is the source flux integrated within the FWHM boundary of the best-fit elliptical Gaussian, $D$ is the distance to the source, $\kappa_{\nu}$ is the dust opacity, and $B_{\nu}(T_{dust})$ is the Planck function at the dust temperature $T_{dust}$. We assumed $\mathbb{R}=100$ and $\kappa_{1.3 \ \rm{mm}} = 0.9 \ \rm{cm}^2 \ \rm{g}^{-1}$, which is the opacity expected for dust with thin ice mantles at a number density of $10^6$ cm$^{-3}$ \citep{1994A&A...291..943O}. Because the values of $\mathbb{R}$ and $\kappa_{1.3 \ \rm{mm}}$ are uncertain, we assume a $30\%$ uncertainty on our assumed values. We adopted $D=4.59^{+0.38}_{-0.33}$ kpc, the parallax distance to the high-mass star-forming region G23.01-0.41 that also resides within GMC 23.0-0.4 \citep{2009ApJ...693..424B}. To estimate the dust temperature, we analyzed \textit{Herschel} sub-mm data \citep{2010PASP..122..314M} using a single temperature graybody model and the methods and assumptions described by \citet{2015ApJ...815..130G}. We calculated dust temperatures in the range of $T_{dust} \sim16 - 20$ K and adopted $T_{dust}=18 \pm 2$ K, though the error may be larger because the dust temperature was derived from data at a larger angular scale ($\sim 35''$). We measured $F_{\nu}$ within the FWHM boundary of the best-fit models. Next we determined $M_{vir}$ using an equation given by \citet{1988ApJ...333..821M},
\begin{equation}
M_{vir} = 3\left(\frac{5-2n}{3-n}\right) \frac{R \sigma^2}{G}, 
\end{equation}
 where $n$ specifies the density distribution ($\rho(r) \propto r^{-n}$), $R$ is the radius of the core, $\sigma$ is the line-of-sight velocity dispersion, and G is the gravitational constant. We assumed that $n=1.8$, the average value for the radial density profile index found by \citet{2009ApJ...696..268Z} in IRDC cores. We estimated $R$ from the geometric mean of the best-fit core radii $R = \sqrt{R_{1} R_{2}}$ and measured $\sigma$ from the SMA $^{13}$CO(2-1) spectra toward the best-fit locations of the cores. We also detected C$^{18}$O(2-1) toward MM2 and MM3 and measured $\sigma$(C$^{18}$O) $\approx \sigma$($^{13}$CO). Because the continuum data provide no kinematic information, the association between the continuum emission from a molecular core and a particular velocity component is sometimes ambiguous. Although we detected no obvious compact molecular line emission associated solely with the continuum cores, the SMA $^{13}$CO emission peaking at $V_{\rm{LSR}} \sim 77$ km s$^{-1}$ suggests that the cores are associated with the pre-shock velocity component. Consequently, we have measured $\sigma$ from the bright velocity component at $V_{\rm{LSR}} \sim 77$ km s$^{-1}$. We display the results of our virial analysis in Table~\ref{tab:cont}. Although we have only determined upper limits on the virial parameter for MM1 and MM2 due to the unresolved core radii, the upper limit radii are similar to the expected size of pre-stellar molecular cores \citep{1999MNRAS.305..143W}, so the true virial parameters are unlikely to be much smaller than our upper limit estimates. Our analysis indicates that all of the cores embedded within G23.33-0.30 have $\alpha > 1$, implying that they are not currently unstable to collapse. 
 

\begin{deluxetable}{cccc}

\tabletypesize{\small}
\tablecolumns{4}
\tablecaption{Molecular Core Properties \label{tab:cont}}

\tablehead{\colhead{}&
		 \colhead{MM1}&
		 \colhead{MM2}&
	         \colhead{MM3}}
\startdata
$l$ (deg) & 23.32384$\pm$0.00002 & 23.32254$\pm$0.00003 & 23.31680$\pm$0.00003 \\
$b$ (deg) & -0.30314$\pm$0.00001 & -0.30459$\pm$0.00002 & -0.30465$\pm$0.00002 \\
$R_1$ (arcsec) & $2.3\pm0.1$ & $2.3\pm0.2$ & $3.2\pm0.3$  \\
$R_2$ (arcsec) & $<1.1$ & $<1.1$ & $1.2\pm0.1$ \\
$R$ (pc) & $<0.036$ & $<0.035$  & $0.044\pm0.004$  \\
$F_{\nu}$ (mJy) & 19.1$\pm$0.2 & 13.9$\pm$0.2 & 13.0$\pm$0.3 \\
$M$ (M$_{\odot}$) & $9.9\pm4.7$ & $7.2\pm3.4$ & $6.7\pm3.2$ \\
$\sigma$($^{13}$CO) (km s$^{-1}$) & $0.57\pm0.07$ & $0.76\pm0.02$ & $0.43\pm0.03$ \\
$\sigma$(C$^{18}$O) (km s$^{-1}$) & \nodata & $0.83\pm0.09$ & $0.46\pm0.06$ \\
$M_{vir}$ (M$_{\odot}$) & $<36$ & $<66$ & $26\pm3$ \\
$\alpha$ & $<3.7$ & $<9.2$ & $3.9\pm1.9$ \\
\enddata
\end{deluxetable}

We also investigated the stability of the larger-scale filament by comparing G23.33-0.30's mass to the mass expected for a collapsing filament. We estimated G23.33-0.30's mass using \textit{Herschel} sub-mm data and the methods and assumptions described in \citet{2015ApJ...815..130G}. We calculated a gas mass of $M\sim600$~M$_{\odot}$ for the $\sim 3\times0.2$ pc portion of G23.33-0.30 visible in Figure~\ref{fig:MIR_VLA_masers}, which corresponds to a linear mass density of $M/l \sim 200$ $M_{\odot}$ pc$^{-1}$. For a typical dense molecular clump (pc size scale), the minimum mass required to form an 8 $M_{\odot}$ star is $\sim260$ M$_{\odot}$ \citep{2017ApJ...841...97S}, assuming a star formation efficiency of 30\% and a \citet{2001MNRAS.322..231K} initial mass function. Although this places G23.33-0.30 in the category of potentially high-mass star-forming filaments, the filament's highly blueshifted and turbulent gas make this less certain. The critical linear mass density above which a molecular filament is unstable to collapse is given by $(M/l)_{crit} = 84(\Delta V)^2 \ M_{\odot}$ pc$^{-1}$ \citep{2010ApJ...719L.185J}, where $\Delta V$ is the FWHM line width. Although the velocity dispersion varies throughout G23.33-0.30, a typical value is $\sigma \sim 2$ km s$^{-1}$, which corresponds to $\Delta V \sim 5$ km s$^{-1}$. This value provides a critical linear mass density of $(M/l)_{crit} \sim 2000$ $M_{\odot}$ pc$^{-1}$, much larger than our estimate of G23.33-0.30's linear mass density. Thus, like G23.33-0.30's embedded cores, the filament is not massive enough to collapse given the turbulent gas.

\section{Discussion}
\label{sec:disc}

\subsection{Evidence of a Large-Scale Shock}

Although NH$_3$(3,3) maser emission typically indicates shocked gas, the positions and velocities of the masers alone cannot distinguish between the protostellar outflow and SNR-MC interaction scenarios. On the other hand, the sharp velocity discontinuity across the width of the filament, which corresponds to an increase in temperature and velocity dispersion, implies a large-scale shock and greatly favors a SNR-MC interaction scenario. Moreover, the SiO(5-4) emission, which traces shocked gas \citep{1997A&A...322..296C}, is extended throughout the filament (Fig.~\ref{fig:NH3_SiO_CS_kin}). While shocks from protostellar outflow could be consistent with the NH$_3$(3,3) maser emission, the resulting SiO emission would be confined to narrow outflow jets emanating from the continuum cores. Since protostellar outflows cannot account for such extended SiO emission, it is more likely that a large-scale shock from a SNR is responsible. 

\begin{figure}[h]
\centering
\includegraphics[scale=1]{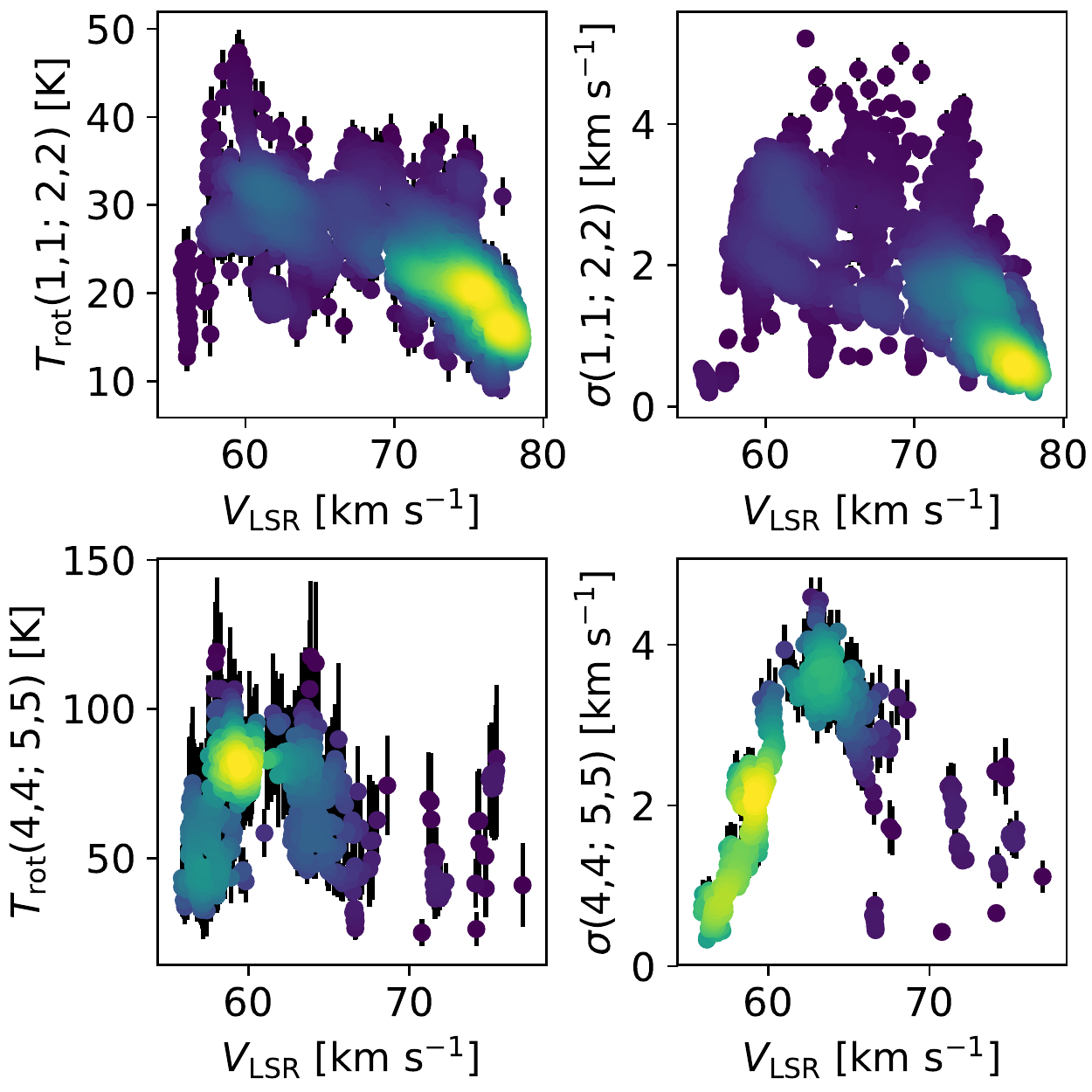}
\caption{Plots of the NH$_3$ model best-fit values of $T_{\rm{rot}}$ and $\sigma$ vs. $V_{\rm{LSR}}$, where the symbol color corresponds to the density of points. The upper plots display the parameter values derived from the NH$_3$(1,1) and (2,2) spectra, while the lower plots show the values derived from the NH$_3$(4,4) and (5,5) spectra. We show only the fit results that have parameter errors below the 75$^{\rm{th}}$ percentile and parameter values that are neither pegged to their maximum nor minimum values.}
\label{fig:NH3_pars}
\end{figure}

Figure~\ref{fig:NH3_pars} shows plots of $T_{\rm{rot}}$(1,1; 2,2) and $\sigma$(1,1; 2,2) vs. $V_{\rm{LSR}}$ and $T_{\rm{rot}}$(4,4; 5,5) and $\sigma$(4,4; 5,5) vs. $V_{\rm{LSR}}$. The NH$_3$(4,4) and (5,5) emission is clearly more sensitive to the hot gas component than the NH$_3$(1,1) and (2,2) emission. The plots of $T_{\rm{rot}}$(1,1; 2,2) and $\sigma$(1,1; 2,2) vs. $V_{\rm{LSR}}$ show that the gas at the pre-shock velocity ($V_{\rm{LSR}}=77-78$ km s$^{-1}$) is generally colder and has a lower velocity dispersion than the turbulent component at lower $V_{\rm{LSR}}$, but the correlations among the parameters is not particularly tight. On the other hand, $T_{\rm{rot}}$(4,4; 5,5) and $\sigma$(4,4; 5,5), which are sensitive to the shocked component, exhibit a more coherent relationship with $V_{\rm{LSR}}$. The emission at $V_{\rm{LSR}} < 60$ km s$^{-1}$ displays positive correlations among $T_{\rm{rot}}$(4,4; 5,5), $\sigma$(4,4; 5,5), and $V_{\rm{LSR}}$, while the emission at $V_{\rm{LSR}} > 60$ km s$^{-1}$ displays negative correlations among these parameters. We speculate that these trends are a signature of the impact that triggered the NH$_3$(3,3) maser emission.

In G23.33-0.30's reference frame ($V_{\rm{LSR}} \sim 77$ km s$^{-1}$), the gas component associated with the NH$_3$(3,3) masers at $V_{\rm{LSR}} \sim 56$ km s$^{-1}$ approaches the filament with a relative velocity of at least $ 20$ km s$^{-1}$. The impacting gas component forms a shock and excites maser emission as it interacts with the filament. The impulse of the shock accelerates the filament gas, blueshifting it to lower $V_{\rm{LSR}}$, as well as increasing its temperature and velocity dispersion. This shock acceleration of the filament gas accounts for the negative correlations among $T_{\rm{rot}}$(4,4; 5,5), $\sigma$(4,4; 5,5), and $V_{\rm{LSR}}$ for $V_{\rm{LSR}} > 60$ km s$^{-1}$. Simultaneously, the impacting gas component is decelerated by the interaction with the filament due to G23.33-0.30's inertia. In the reference frame of the component at $V_{\rm{LSR}} \sim 56$ km s$^{-1}$, it encounters the dense filament approaching quickly and is impacted, redshifting it to higher $V_{\rm{LSR}}$, as well as increasing its temperature and velocity dispersion. This deceleration of the impacting component due to G23.33-0.30's inertia accounts for the positive correlations among $T_{\rm{rot}}$(4,4; 5,5), $\sigma$(4,4; 5,5), and $V_{\rm{LSR}}$ for $V_{\rm{LSR}} < 60$ km s$^{-1}$. Consequently, the hot, highly turbulent component at $V_{\rm{LSR}} = 60-68$ km s$^{-1}$ likely represents the turbulent wake of the shock \citep{2016MNRAS.457.4470P}.

\begin{figure}[h]
\centering
\includegraphics[scale=1]{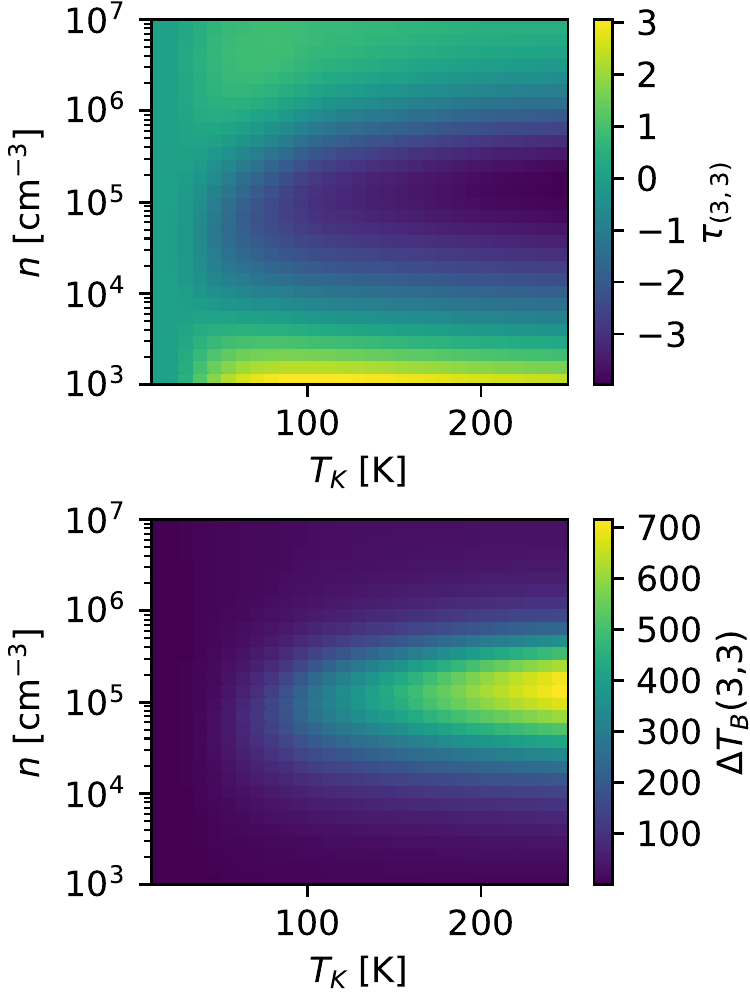}
\caption{Plot of the RADEX NH$_3$(3,3) optical depth ($\tau_{(3,3)}$) (top) and brightness temperature ($\Delta T_B$(3,3)) (bottom) as a function of the kinetic temperature ($T_k$) and number density ($n$) for a column density of $N=10^{15.5}$ cm$^{-2}$ and a velocity dispersion of $\sigma=1.7$ km s$^{-1}$.}
\label{fig:radex_maser}
\end{figure}

Assuming the maser at $V_{\rm{LSR}} = 76.4$ km s$^{-1}$ represents genuine nonthermal emission, it likely signals another shock front within the filament. This source is associated with gas at higher $V_{\rm{LSR}}$ and is much fainter than the other NH$_3$(3,3) masers. Figure~\ref{fig:dec-vlsr} suggests that the maser, MM1, and MM2 are all associated with a section of the filament that exhibits values of $T_{\rm{rot}}$, $\sigma$, and $V_{\rm{LSR}}$ that are intermediate between the turbulent component and the pre-shock component. Given that this portion of the filament is situated between a highly shocked region and an unshocked region of the filament, we speculate that it is at an earlier stage of shock interaction than the hot, turbulent component. The moderate temperature of this intermediately shocked component may in part explain the faintness of the maser at $V_{\rm{LSR}} = 76.4$ km s$^{-1}$. Figure~\ref{fig:radex_maser} shows a non-LTE RADEX \citep{2007A&A...468..627V} plot of the expected NH$_3$(3,3) optical depth ($\tau_{(3,3)}$) and brightness temperature ($\Delta T_B$(3,3)) as a function of the gas kinetic temperature ($T_k$) and number density ($n$) for the derived beam-averaged column density ($N_{\rm{NH_3}}=10^{15.5}$ cm$^{-2}$) and velocity dispersion ($\sigma=1.7$ km s$^{-1}$) of the intermediately shocked component. Large brightness temperatures and negative optical depths, which indicate strong masing, are only achieved for larger temperatures. At low temperatures, the RADEX model predicts that the gas should be either weakly masing or non-masing. Given the low rotational temperature of the intermediately shocked component ($T_{\rm{rot}}$(1,1; 2,2)$\sim 20-25$ K), the lower $\Delta T_B$(3,3) of the maser is expected. If the shock continues to heat this section of the filament, it is possible that the maser's brightness temperature will increase. 

\begin{figure}[h]
\centering
\includegraphics[scale=1]{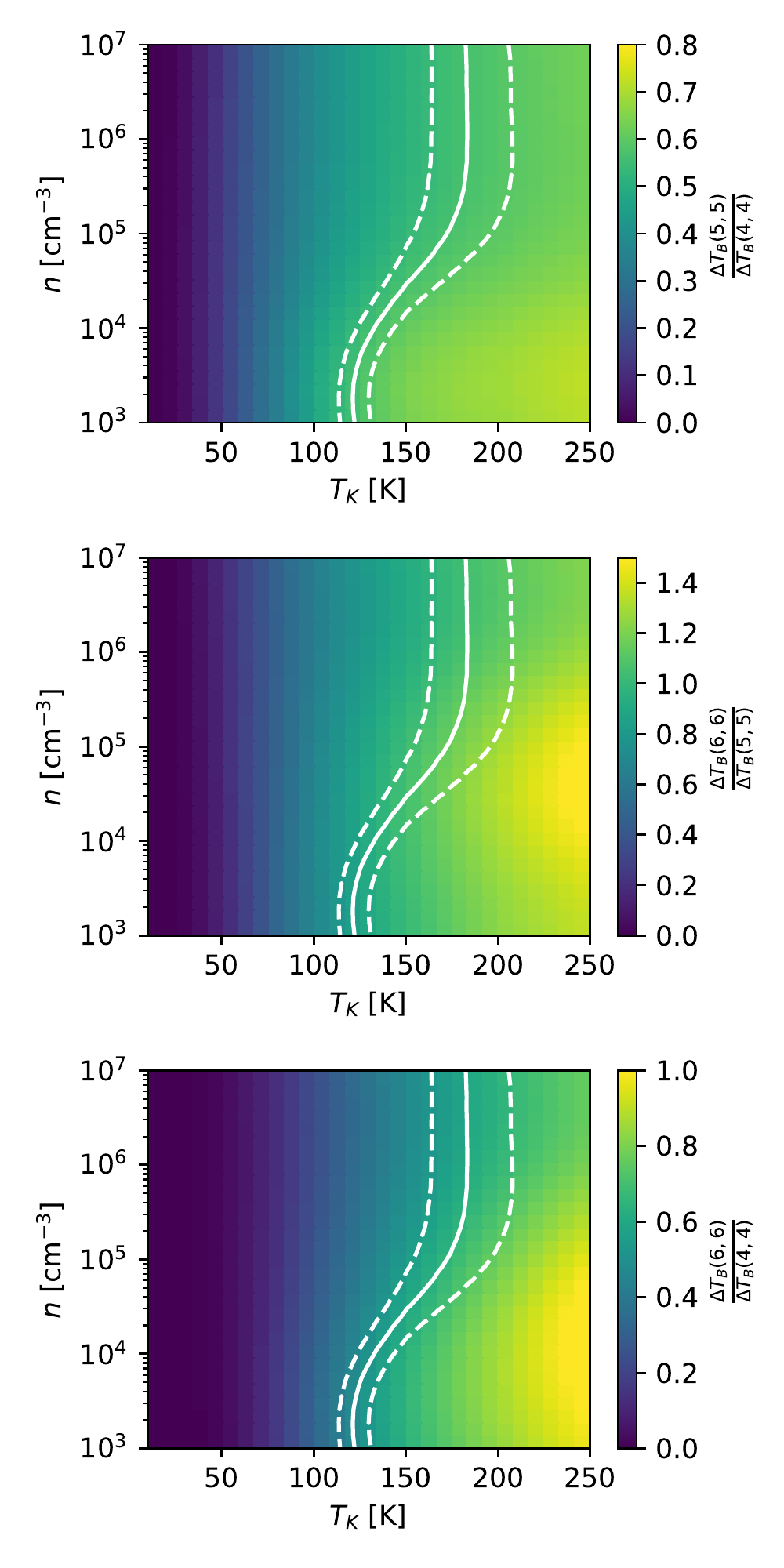}
\caption{Plot of RADEX brightness temperature ratios as a function of the kinetic temperature ($T_k$) and number density ($n$): $\frac{\Delta T_B(5,5)}{\Delta T_B(4,4)}$ (top), $\frac{\Delta T_B(6,6)}{\Delta T_B(5,5)}$ (middle), and $\frac{\Delta T_B(6,6)}{\Delta T_B(4,4)}$ (bottom). We assume a column density of $N=10^{15.5}$ cm$^{-2}$ and a velocity dispersion of $\sigma=1.75$ km s$^{-1}$. The solid white line in each plot marks the temperatures and densities corresponding to the $\frac{\Delta T_B(5,5)}{\Delta T_B(4,4)} = 0.570 \pm 0.026$ measured towards the peak of the NH$_3$(6,6) emission. The dashed lines mark the 1-$\sigma$ uncertainties on the measured amplitude ratio.}
\label{fig:radex_ortho}
\end{figure}

The large rotational temperatures of the hot, turbulent component demonstrate that the shock has heated the gas in the filament, but the NH$_3$(4,4) and (6,6) data also provide evidence of dust heating. NH$_3$ modeling by \citet{2013ApJ...770L...2F} suggests that ortho-NH$_3$ ($K = 3n$) forms preferentially over para-NH$_3$ ($K \neq 3n$) on the surfaces of cold ($T < 30$ K) dust grains. In addition, the ortho-NH$_3$ ground state is at a lower energy than the para-NH$_3$ ground state, resulting in a larger amount of energy needed to desorb para-NH$_3$ than ortho-NH$_3$ \citep{1999ApJ...525L.105U}. Consequently, shocks that heat the icy mantles of cold dust grains release ortho-enhanced NH$_3$ into the gas phase and result in an ortho-to-para abundance ratio (OPR) larger than the statistical equilibrium value of OPR $=1$. Enhanced OPRs have previously been observed by \citet{1999ApJ...525L.105U} in the L1157 outflow and by \citet{2017MNRAS.468.2093D} toward SNR-MC interactions. To confirm the presence of ortho-enhanced gas associated with G23.33-0.30, we fit the NH$_3$(4,4) through (6,6) spectra toward the peak of the NH$_3$(6,6) emission. We calculated brightness temperature ratios of $\frac{\Delta T_B(5,5)}{\Delta T_B(4,4)} = 0.57 \pm 0.03$, $\frac{\Delta T_B(6,6)}{\Delta T_B(5,5)} = 2.71 \pm 0.13$, and $\frac{\Delta T_B(6,6)}{\Delta T_B(4,4)} = 1.54 \pm 0.06$. Figure~\ref{fig:radex_ortho} shows the NH$_3$(4,4) through (6,6) brightness temperature ratios predicted by RADEX for a range of temperatures and densities, assuming the column density ($N=10^{15.5}$ cm$^{-2}$) and velocity dispersion ($\sigma=1.75$ km s$^{-1}$) measured from the NH$_3$(4,4) and (5,5) spectra and OPR$=1$. The temperatures and densities corresponding to our measured value of $\frac{\Delta T_B(5,5)}{\Delta T_B(4,4)}$ are indicated by the white lines overlaid on each plot. If the turbulent gas component had OPR=1, the lines would intersect our measured values of $\frac{\Delta T_B(6,6)}{\Delta T_B(5,5)}$ and $\frac{\Delta T_B(6,6)}{\Delta T_B(4,4)}$. However, our measured values for $\frac{\Delta T_B(6,6)}{\Delta T_B(5,5)}$ and $\frac{\Delta T_B(6,6)}{\Delta T_B(4,4)}$ are larger than any of the values on either plot, suggesting that OPR $>1$ for the turbulent gas component. Although masing in the NH$_3$(6,6) transition offers another possible explanation for the large ortho-to-para brightness ratios, the large velocity dispersion of the NH$_3$(6,6) line ($\sigma \sim 1.5$ km s$^{-1}$) make this unlikely. Thus, G23.33-0.30 is associated with ortho-enhanced gas, implying that the shock has heated the filament's cold dust and sublimated NH$_3$ from their icy mantles. 

\subsection{W41's SNR-MC Interaction with GMC G23.0-0.4}

Although molecular clouds can experience shocks due to protostellar jets or \Hii\ regions, supernova shocks deliver a stronger impulse over a much shorter time span. Supernovae release roughly $10^{51}$ ergs of energy nearly instantaneously, sending powerful shock waves over tens of parsecs. Despite the fact that only $\sim5-10\%$ of the total energy is converted into kinetic energy in the shock \citep{2015MNRAS.451.2757W}, the energy in a supernova shock can be sufficient to disrupt and disperse molecular clouds and cores. Using a simple $E_{shock} = \frac{1}{2} M_{shocked} \Delta V^2$ energy transfer analysis, the mass displaced by a SNR shock is approximately given by $M_{shocked} \sim \frac{2 \Omega f_{kin} E_{SN}}{4 \pi \Delta V^2}$, where $\Omega$ is the molecular cloud's solid angle at its distance from the supernova, $f_{kin}$ is the kinetic efficiency of the SNR shock, $E_{SN}$ is the total energy of the supernova, and $\Delta V^2$ is the change in velocity of the shocked gas. Assuming that $E_{SN} = 10^{51}$ ergs, $f_{kin} = 5\%$, $\Delta V = 10$ km s$^{-1}$, and $\Omega$ is the solid angle of a spherical cloud with a radius of 0.5 pc radius at 5, 10, or 20 pc away from the supernova, the gas mass displaced by the shock is 130, 35, or 9 M$_{\odot}$, respectively.

Given the uncertainty in the shock properties, the mass in the shocked portion of G23.33-0.30 is comparable to the mass able to be displaced by a SNR shock, implying that a SNR is a plausible source for the shock that is accelerating the molecular filament gas.  Considering the suggestions of a SNR-MC interaction between SNR W41 and GMC G23.0-0.4 \citep{2013ApJ...773L..19F}, W41 is an attractive progenitor for the large-scale shock impacting G23.33-0.30. A SNR shock would likely supply enough energy to explain G23.33-0.30's highly blueshifted emission, broad turbulent line widths, and increased temperature and velocity dispersion. Thus, we speculate that the turbulent, blueshifted gas observed in G23.33-0.30 is the result of a large scale shock originating from the nearby SNR W41. 

\begin{figure}[h]
\centering
\includegraphics[scale=0.75]{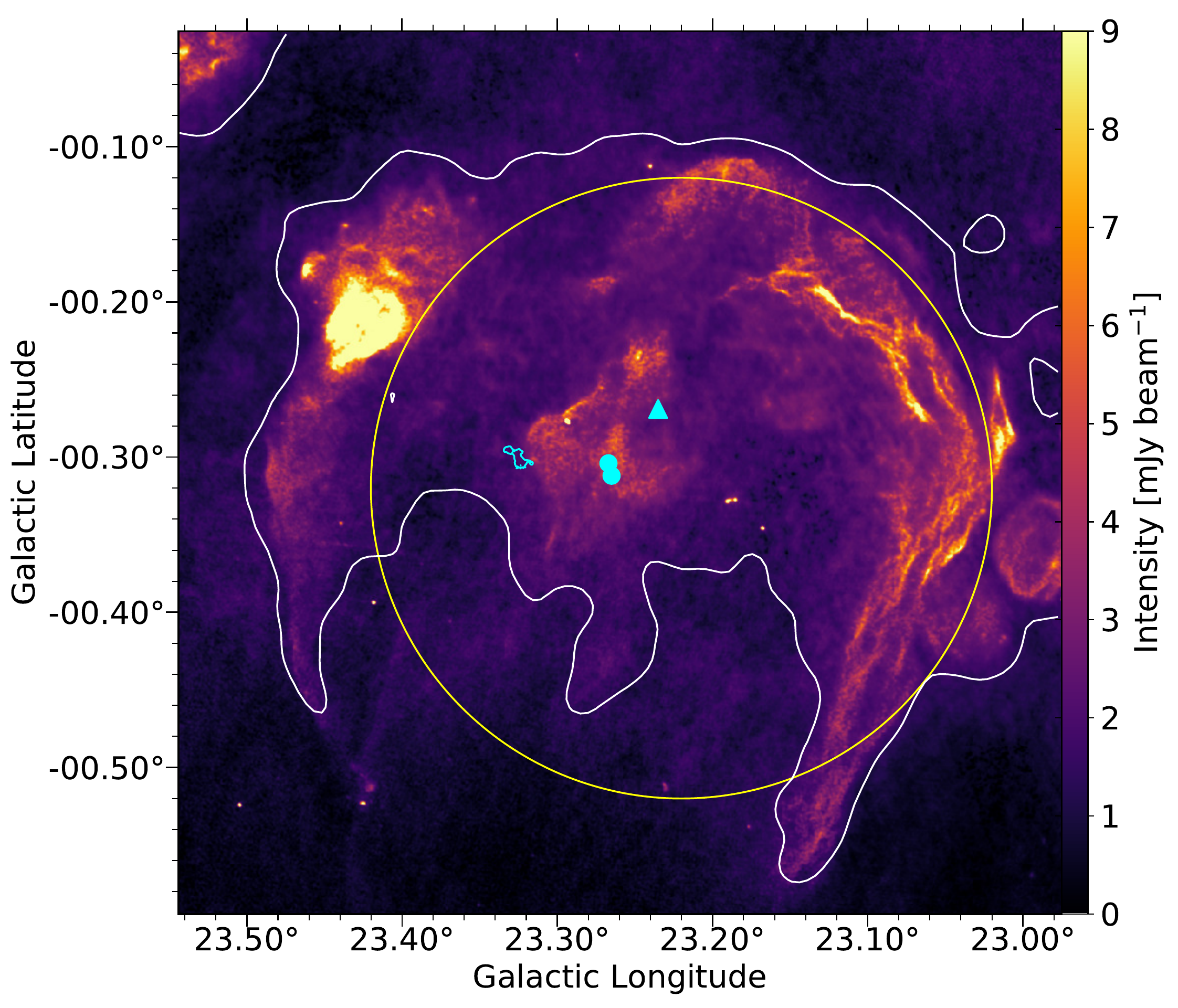}
\caption{Color shows the MAGPIS 20 cm continuum data. The white contours show the VGPS 21 cm continuum data at 30 K and the cyan contours show the VLA NH$_3$(2,2) integrated intensity data from G23.33-0.30 at 50 mJy beam$^{-1}$ km s$^{-1}$. The cyan circles coincident with the central 20 cm continuum peak indicate the positions of two 1720 MHz OH maser candidates, which have velocities near 75 km s$^{-1}$, and the cyan triangle shows the position of the pulsar candidate CXOU J183434.9-08444. The large yellow circle displays the fitted position and FWHM size of the extended component of HESS J1834-087's TeV $\gamma$-ray emission.}
\label{fig:W41_20cm}
\end{figure}

Figure~\ref{fig:W41_20cm} shows the Multi-Array Galactic Plane Imaging Survey \citep[MAGPIS;][]{2006AJ....131.2525H} 20 cm continuum emission from W41 in color and the VLA Galactic Plane Survey \citep[VGPS;][]{2006AJ....132.1158S} 21 cm data as contours. W41 is an asymmetric shell-type supernova \citep{1991PASP..103..209G,1992AJ....103..943K} suspected of interacting with the nearby GMC G23.0-0.4 at $V_{\rm{LSR}} = 77$ km s$^{-1}$ \citep{2015ApJ...811..134S}. \citet{2013ApJ...773L..19F} detected two 1720 MHz OH maser candidates, known to trace SNR-MC interactions \citep{2002Sci...296.2350W}, coincident with W41's central continuum peak. W41 is also coincident with HESS J1834-087 \citep{2006ApJ...636..777A,2006ApJ...638L.101A}, a source of TeV emission thought to be triggered by W41's interaction with a GMC \citep{2007ApJ...657L..25T}. Deep follow-up observations with H.E.S.S. revealed that the TeV emission is composed of a point-like source and an extended component \citep{2015A&A...574A..27H}. Although the pulsar candidate CXOU J183434.9-08444 \citep{2011ApJ...735...33M} may account for the point-like component of the TeV emission, \citet{2015A&A...574A..27H} argued that the extended TeV emission is best explained by the SNR-MC interaction.

\begin{figure}[h]
\centering
\includegraphics[scale=0.75]{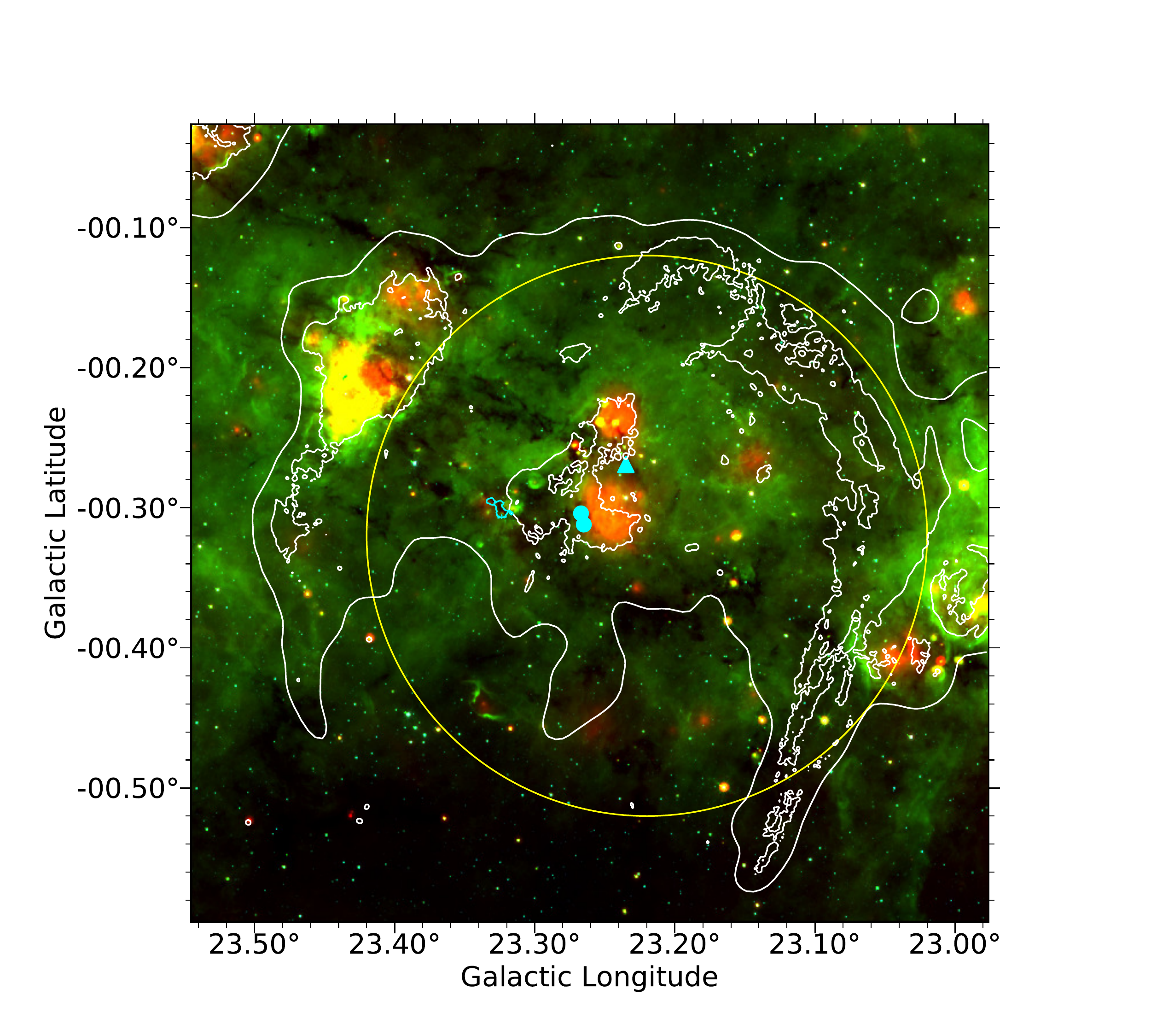}
\caption{Color shows the MIPSGAL 24 $\mu$m (red), GLIMPSE 8 $\mu$m (green) and 3.6 $\mu$m (blue) MIR data. For reference we show the MAGPIS 20 cm continuum data with contours at 3 mJy beam$^{-1}$, as well as the data overlaid in Figure~\ref{fig:W41_20cm}.}
\label{fig:W41_MIR}
\end{figure}

\begin{figure}[h]
\centering
\includegraphics[scale=0.33]{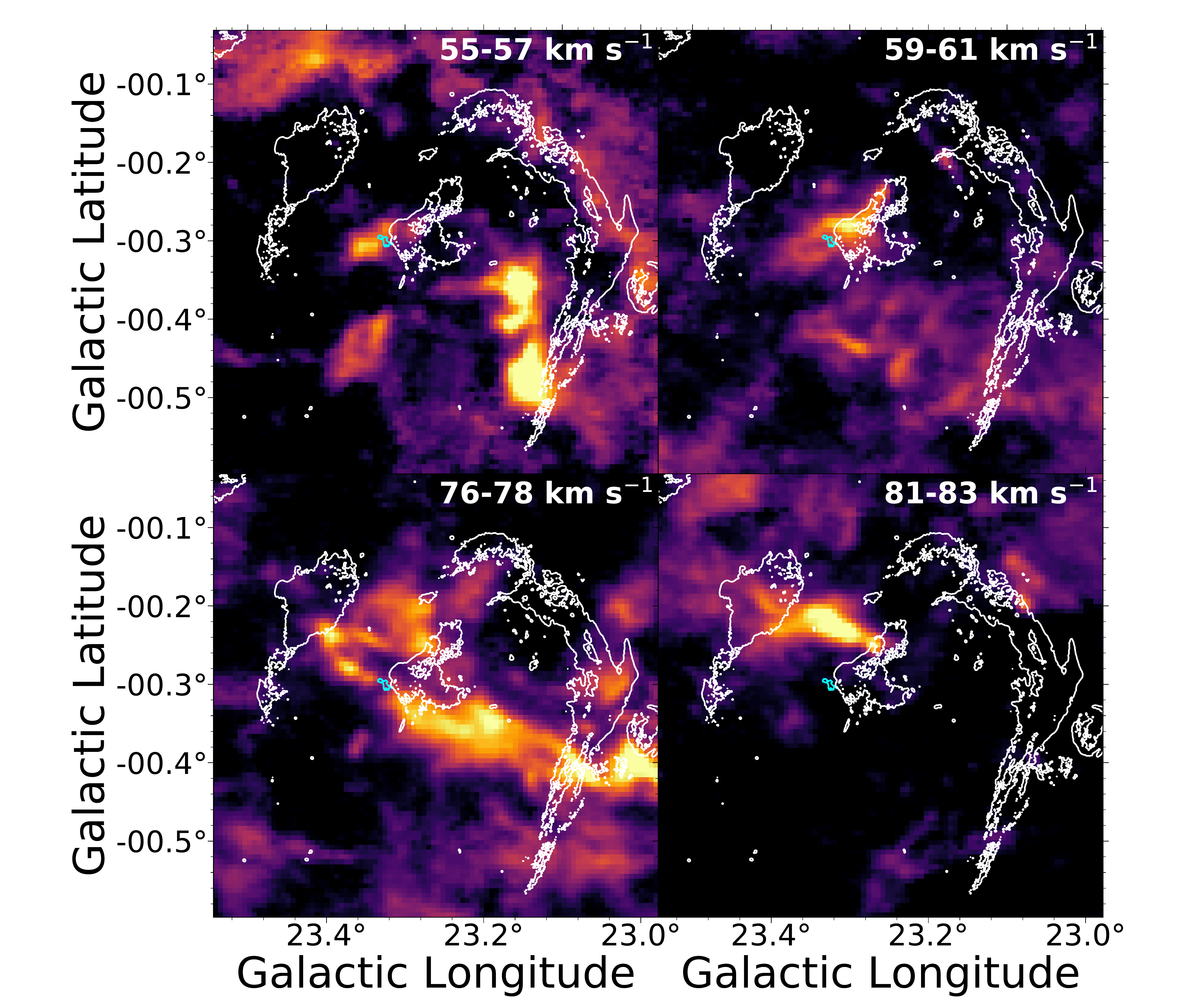}
\caption{Color shows the GRS $^{13}$CO(1-0) integrated intensity over the ranges specified in each panel. We show the MAGPIS 20 cm continuum data with contours at 3 mJy beam$^{-1}$ for reference. The cyan contours show the VLA NH$_3$(2,2) integrated intensity data from G23.33-0.30 at 50 mJy beam$^{-1}$ km s$^{-1}$.}
\label{fig:W41_GRS}
\end{figure}

\begin{figure}[h]
\centering
\includegraphics[scale=0.75]{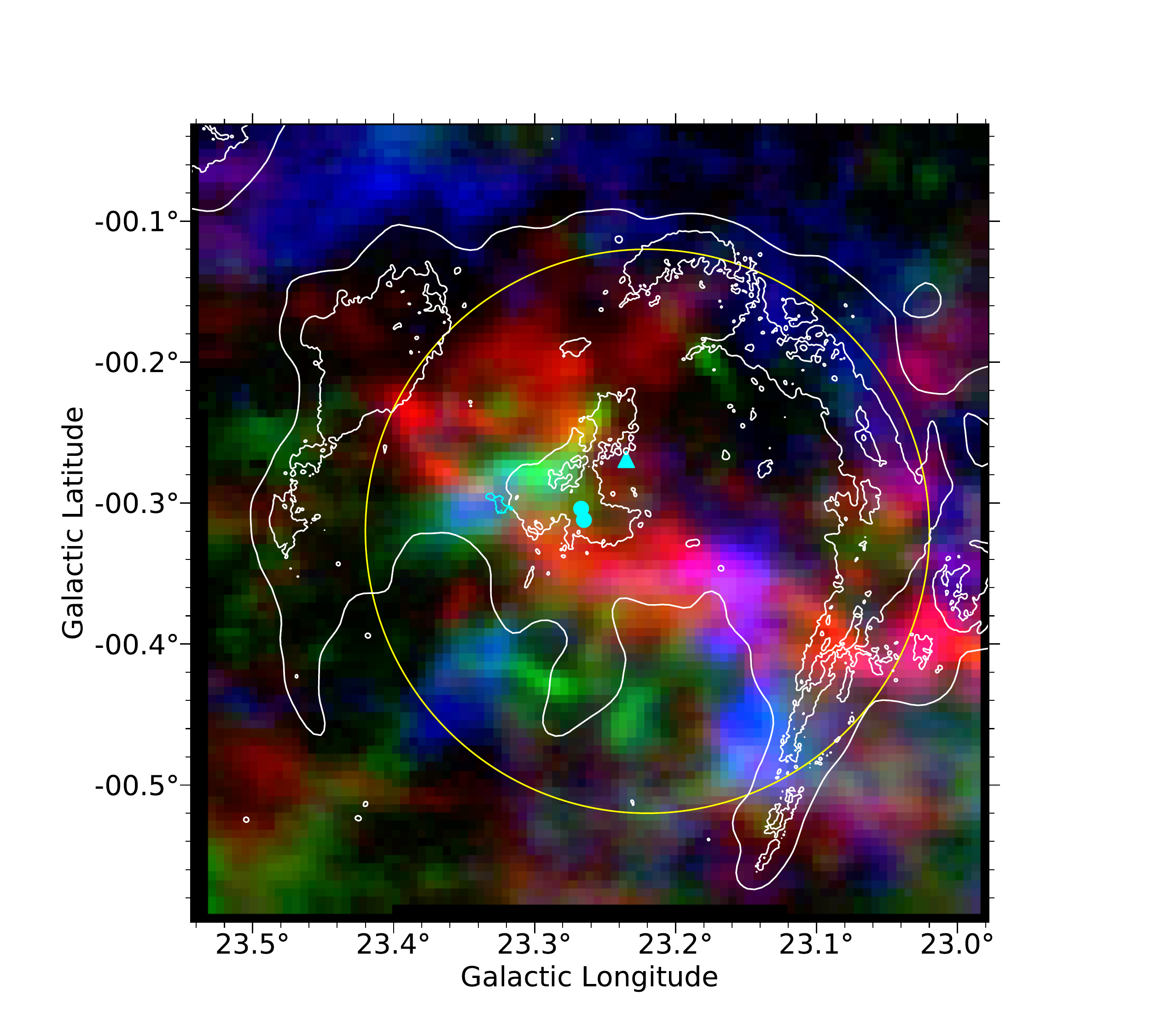}
\caption{Color shows the GRS $^{13}$CO integrated intensity over the range $55-57$ km s$^{-1}$ (blue), $59-61$ km s$^{-1}$ (green), and $76-78$ km s$^{-1}$ (red). Symbols and contours as in Figure~\ref{fig:W41_MIR}.}
\label{fig:W41_GRS_RGB}
\end{figure}

The MIR emission toward the W41 region from GLIMPSE and the MIPS Galactic Plane Survey \citep[MIPSGAL; ][]{2009PASP..121...76C} is shown in Figure~\ref{fig:W41_MIR}, displaying a rich and complicated star-forming complex. These data reveal filamentary IRDCs and several sources of bright MIR emission, but provide no kinematic information about the molecular gas near W41. In order to better understand the kinematics of the molecular gas in the W41 region, we analyzed the Galactic Ring Survey \citep[GRS; ][]{2006ApJS..163..145J} $^{13}$CO(1-0) data. The VLA NH$_3$ data revealed three prominent velocity components: a narrow component at $V_{\rm{LSR}}\sim56$ km s$^{-1}$ that appears to be associated with three of the NH$_3$(3,3) masers, a broad, turbulent component at $V_{\rm{LSR}}\sim60$ km s$^{-1}$, and a relatively asymmetric line peaked at $V_{\rm{LSR}}\sim77$ km s$^{-1}$, which is associated with the NH$_3$(3,3) maser at $V_{\rm{LSR}}\sim76$ km s$^{-1}$. We inspected the $^{13}$CO data for emission associated with these velocity components and created maps of the integrated intensity in 2 km s$^{-1}$ windows centered on these velocities. We also found another distinct $^{13}$CO(1-0) velocity component peaking at $81-82$ km s$^{-1}$, which is spatially and spectrally adjacent to the other emission. Figure~\ref{fig:W41_GRS} shows the GRS $^{13}$CO(1-0) data integrated in the velocity ranges indicated in each panel, with the MAGPIS continuum overlaid for comparison. The lower velocity emission centered at 56 and 60 km s$^{-1}$ displays emission near the positions of W41's central 20 cm continuum peak and G23.33-0.30. On the other hand, the emission at 77 km s$^{-1}$ clearly traces GMC G23.0-0.4, but exhibits a conspicuous deficit in emission where W41's central continuum emission peaks. Finally, the 82 km s$^{-1}$ component traces a MC, seen as a collection of IRDC filaments in Figure~\ref{fig:W41_MIR}, that could represent a nearby background cloud or some other component of the GMC. \citet{2007ApJ...657L..25T} also noted the association between the lower velocity emission, W41's continuum, and HESS J1834-087, but they assumed that the emission represented a separate GMC with which W41 was interacting. If this were true, the emission from GMC G23.0-0.4 would likely be uncorrelated with the central 20 cm continuum peak, rather than anti-correlated. Moreover, Figure~\ref{fig:W41_GRS_RGB} shows that the lower velocity emission at 56 and 60 km s$^{-1}$ and the GMC emission at 77 km s$^{-1}$ also appear to be anti-correlated. In light of the agreement between the velocities of the 1720 MHz OH maser candidates and the GMC, as well as the apparent anti-correlation between the lower velocity emission and the GMC emission, we argue that these lower velocity components are associated with GMC G23.0-0.4 and the SNR-MC interaction. 

\citet{2013ApJ...773L..19F} also detected an OH absorption feature at 76 km s$^{-1}$, which places W41 within or behind G23.0-0.4. In addition, \citet{2008AJ....135..167L} measured a maximum \Hi\ absorption velocity of $78 \pm 2$ km s$^{-1}$ toward W41, indicating that W41's progenitor may have formed within G23.0-0.4. The \citet{2014ApJ...783..130R} near kinematic distance for $V_{\rm{LSR}} = 77$ km s$^{-1}$ toward G23.33-0.30 is $4.41^{+0.27}_{-0.26}$ kpc and the maser parallax distance to the nearby HMSFR G23.01-0.41 is $4.59^{+0.38}_{-0.33}$ kpc \citep{2009ApJ...693..424B}. Considering that the GRS $^{13}$CO(1-0) data strongly imply that G23.01-0.41 and G23.33-0.30 both reside within GMC G23.0-0.4, we adopt the maser parallax distance for W41. W41's angular size of $\sim0\fdg5$ implies that the SNR has a physical diameter of $\sim40$ pc. This size is in agreement with estimates from the Sedov relation \citep{1959JApMM..23..568S}, assuming a reasonable average density ($\sim 6$ cm$^{-3}$) and SNR age ($\sim 2 \times 10^{5}$ yr) \citep{2007ApJ...657L..25T}.

Given that these data are consistent with an interaction between SNR W41 and GMC G23.0-0.4, we further speculate that the $^{13}$CO(1-0) emission with $V_{\rm{LSR}}=60-75$ km s$^{-1}$ (not shown) in the vicinity of the W41's central 20 cm continuum peak represents gas from G23.0-0.4 that has been shock-accelerated to its current velocity. The $V_{\rm{LSR}}=56$ km $^{-1}$ component seems to correspond to the component associated with the NH$_3$(3,3) masers in G23.33-0.30. This velocity component, which presumably represents gas moving with the largest velocity relative to G23.33-0.30, appears to be streaming past and interacting with the filament. The existence of unshocked gas within G23.33-0.30, as well as $^{13}$CO(1-0) emission at the pre-shock velocity adjacent to the filament, implies that the interaction between SNR W41 and G23.33-0.30 is ongoing. This agrees with our interpretation of the interferometric data, which exhibits emission at $V_{\rm{LSR}}=77-78$ km s$^{-1}$ (pre-shock gas), $V_{\rm{LSR}}=60-75$ km s$^{-1}$ (turbulent shock wake), and $V_{\rm{LSR}}=56-57$ km s$^{-1}$ (gas entrained in shock front). A caveat to this interpretation is that W41 lies near the plane of the Galaxy, so the line of sight towards the SNR is crowded with molecular clouds at various velocities, which could confuse our interpretation. In addition, feedback from previous generations of high-mass stars can accelerate gas and create molecular gas structures that are physically close, but have different $V_{\rm{LSR}}$. Thus, it is possible that the molecular gas with $V_{\rm{LSR}}=60-75$ km s$^{-1}$ is not currently associated with the GMC, but is in fact a remnant of the larger GMC structure that is being impacted by the SNR shock from W41. On the other hand, it is difficult to imagine such a scenario producing the striking anti-correlation between the GMC velocity component and the $V_{\rm{LSR}}=60$ km s$^{-1}$ velocity component. Although more observations are needed to determine the true relationship between these velocity components, a SNR-MC interaction seems to best explain the large turbulent line widths, heating, NH$_3$(3,3) masers, and blueshifted emission observed in G23.33-0.30.

\begin{figure}[h]
\centering
\includegraphics[scale=0.75]{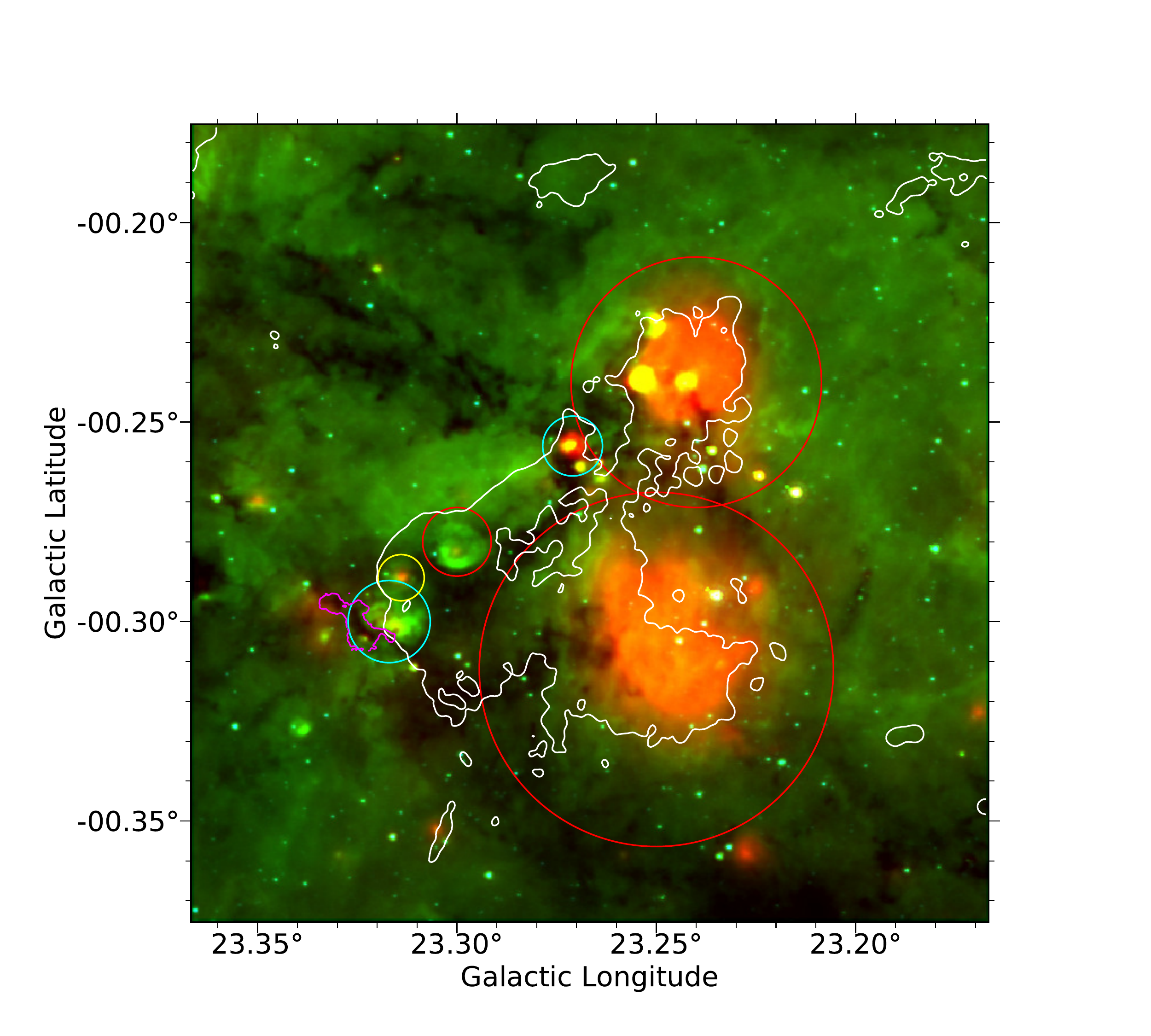}
\caption{Color shows the MIPSGAL 24 $\mu$m (red), GLIMPSE 8 $\mu$m (green) and 3.6 $\mu$m (blue) MIR data. The MAGPIS 20 cm continuum data are shown with white contours at 3 mJy beam$^{-1}$ and the magenta contours show the VLA NH$_3$(2,2) integrated intensity data of G23.33-0.30 at 50 mJy beam$^{-1}$ km s$^{-1}$. The circles show the nearby sources from the \textit{WISE} HII Region Catalog. The red circles indicate known HII regions with RRL detections, the cyan circles indicate candidate HII regions that exhibit continue emission but lack RRL detections, and the yellow circles indicate radio quiet candidate HII regions, which exhibit the MIR characteristics of an HII region but lack radio continue emission and RRL emission.}
\label{fig:W41_cap_MIR}
\end{figure}

An alternative explanation for G23.33-0.30's large turbulent line widths is energy added by nearby \Hii\ regions. Figure~\ref{fig:W41_cap_MIR} shows a zoomed view of the MIR emission toward W41's central continuum peak. The overlaid circles show the positions and sizes of several nearby \Hii\ regions and candidate \Hii\ regions from the \textit{WISE} Catalog of Galactic \Hii\ regions \citep{2014ApJS..212....1A}. While it is clear that these \Hii\ regions cannot account for all of the 20 cm continuum emission toward W41's center, it is likely that they contribute a portion of the emission. The red circles indicate \Hii\ regions that have known velocities from radio recombination lines (RRLs): G023.250-00.240 ($V_{\rm{LSR}} = 76.3$ km s$^{-1}$), G023.265-00.301a ($V_{\rm{LSR}} = 73.1$ km s$^{-1}$), and G023.295-00.280 ($V_{\rm{LSR}} = 61.6$ km s$^{-1}$). Although these velocities may indicate an association with the GMC or the 60 km s$^{-1}$ $^{13}$CO(1-0) component, it is unlikely that these sources could account for the significant energy input implied by G23.33-0.30's large turbulent line widths given their angular separation from G23.33-0.30. The source with the smallest angular separation from G23.33-0.30, the candidate \Hii\ region G023.317-0.300, lacks a reliable velocity from a RRL detection, but its position behind G23.33-0.30 makes it an attractive alternative source for G23.33-0.30's turbulence. On the other hand, our interferometric data exhibit broad SiO(5-4) and CS(5-4) line emission peaking at $V_{\rm{LSR}} = 97$ km s$^{-1}$ near the center of the candidate \Hii\ region, potentially signifying its association with the background source HMSFR G23.44-0.18, which has a maser parallax distance of $5.88^{+1.37}_{-0.93}$ kpc. In addition, the velocity dispersion of G23.33-0.30's turbulent velocity component is $\sigma_{shocked} = 2-5$ km s$^{-1}$, while the velocity dispersions measured towards \Hii\ regions in the RAMPS dataset \citep{2018ApJS..237...27H} are at most $\sigma \lesssim 2$ km s$^{-1}$. Thus, an \Hii\ region would need to input an unusually large amount of energy into G23.33-0.30 to reproduce the measured velocity dispersion. Furthermore, NH$_3$(3,3) masers have previously only been associated with SNR-MC interactions \citep{2016ApJ...826..189M} or energetic outflows from high-mass protostars \citep{1994ApJ...428L..33M,1995ApJ...439L...9K,1995ApJ...450L..63Z}, so NH$_3$(3,3) maser emission resulting from an interaction with an \Hii\ region would also be unusual. Consequently, G23.33-0.30's high levels of turbulence are most likely the result of a SNR-MC interaction.

\subsection{W41's Potential Impact Geometry}

Figure~\ref{fig:W41_GRS} shows that the diameter of W41's shell is much larger than the radial extent of the GMC. If W41's interaction is ongoing and it exploded within G23.0-0.4, then W41's plane of sky diameter must be larger than its size along the line of sight. Dense molecular gas can slow or even stall the expansion of a SNR shock into a MC \citep{1987A&A...184..279T}. Given that MCs can exhibit asymmetric density profiles, core-collapse supernovae shell structures evolving in these environments can also display asymmetries \citep{2009ApJ...706L.106L}. While this is plausible, we might expect W41's expansion out of a dense, massive GMC to sweep up more molecular gas in its shell, whereas this signature is absent in the GRS data. An alternative explanation is that W41's progenitor formed within a MC $\leq 20$ pc away from G23.0-0.4. The 82 km s$^{-1}$ MC shown in Figure~\ref{fig:W41_GRS} could potentially be this cloud. The \citet{2014ApJ...783..130R} near kinematic distance for $V_{\rm{LSR}}=82$ km s$^{-1}$ is $4.60^{+0.26}_{-0.27}$ kpc, consistent with our adopted distance to G23.33-0.30. Although \citet{2008AJ....135..167L} measured a maximum \Hi\ absorption velocity of $78 \pm 2$ km s$^{-1}$ toward W41, this may not preclude W41 from also being associated with a background MC. If W41 expanded out of this background MC, the shock may have blueshifted much of the foreground molecular gas that was previously associated with the background MC, confusing the interpretation. Hence, the uncertainty in the origin point of W41's expansion leaves the impact geometry ambiguous.

We have searched for redshifted gas corresponding to the back side of the expanding supernova shell within G23.0-0.4 to help differentiate between these two scenarios. We detected CS emission peaking at $V_{\rm{LSR}}=98$ and 103 km s$^{-1}$ and SiO emission peaking at $V_{\rm{LSR}}=96$ km s$^{-1}$ near G23.33-0.30, but it is unclear whether this emission is associated with G23.0-0.4 or the interaction. Thus, we examined the GRS $^{13}$CO data for signatures of a redshifted component potentially associated with the higher $V_{\rm{LSR}}$ gas. Although we found emission spanning $V_{\rm{LSR}}=93-106$ km s$^{-1}$ toward the left half of W41, we are cautious to associate this emission with the back side of the W41 shell due to the presence of two background sources. One of these background sources is HMSFR G23.44-0.18, which has a maser parallax distance of $5.88^{+1.37}_{-0.93}$ kpc and $V_{\rm{LSR}}=97\pm3$ km s$^{-1}$, placing it in the Norma arm near the end of the long bar \citep{2009ApJ...693..424B,2014ApJ...783..130R}. The second background source is the \Hii\ region G23.42-0.21, which seems to contribute much of the continuum emission near W41's left edge (Fig.~\ref{fig:W41_MIR}). G23.42-0.21 has a recombination line velocity of 103 km s$^{-1}$ and a maximum \Hi\ absorption velocity of 106 km s$^{-1}$; hence, \citet{2008AJ....135..167L} argued that the absorption velocity was significantly higher than the recombination velocity and that G23.42-0.21 must be assigned to the far distance of $\sim9.9$ kpc. 

\begin{figure}[h]
\centering
\includegraphics[scale=0.7,trim={2cm 4cm 2cm 5cm},clip]{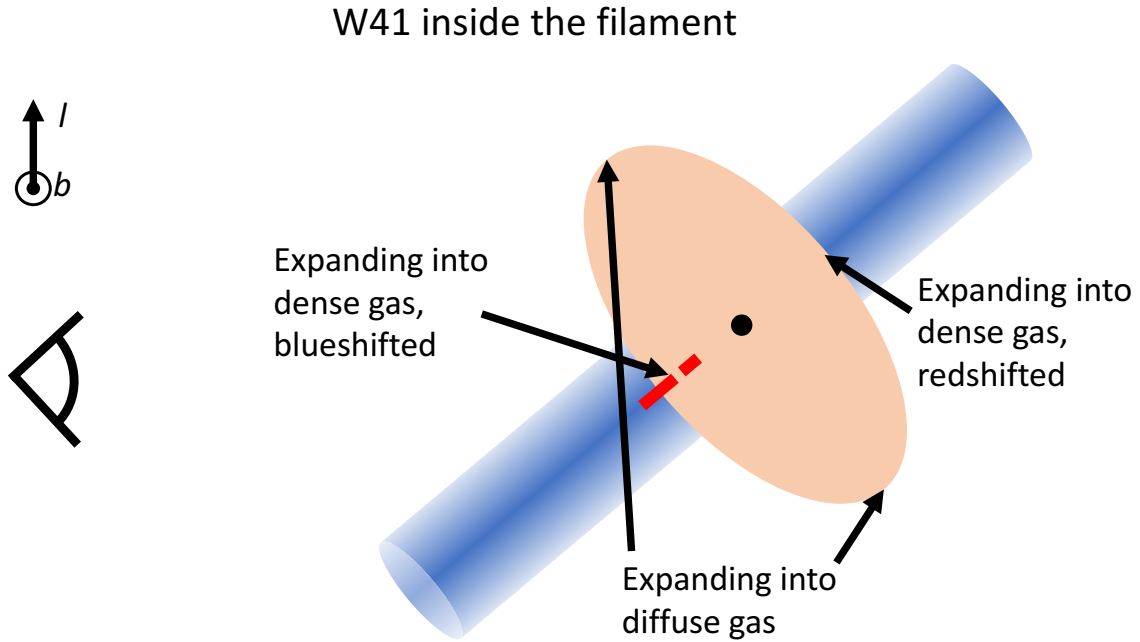}
\caption{Illustration of top-down view of the scenario in which W41 formed near the center of GMC G23.0-0.4. The figure shows the GMC represented as a blue cylinder, W41's shock bubble as a peach ellipse, G23.33-0.30 as red line segments, and the position of W41's progenitor as a black circle. The filament is rotated in the plane of the Galaxy, such that it roughly matches the orientation of the \citet{2014ApJ...783..130R} Scutum arm, in which the GMC may reside \citep{2009ApJ...693..424B}. Although this model does not depend on the exact orientation of the GMC, some degree of rotation would be required in order to detect blueshifted emission at the location of G23.33-0.30.}
\label{fig:cartoon1}
\end{figure}

\begin{figure}[h]
\centering
\includegraphics[scale=0.7,trim={2cm 4cm 2cm 5cm},clip]{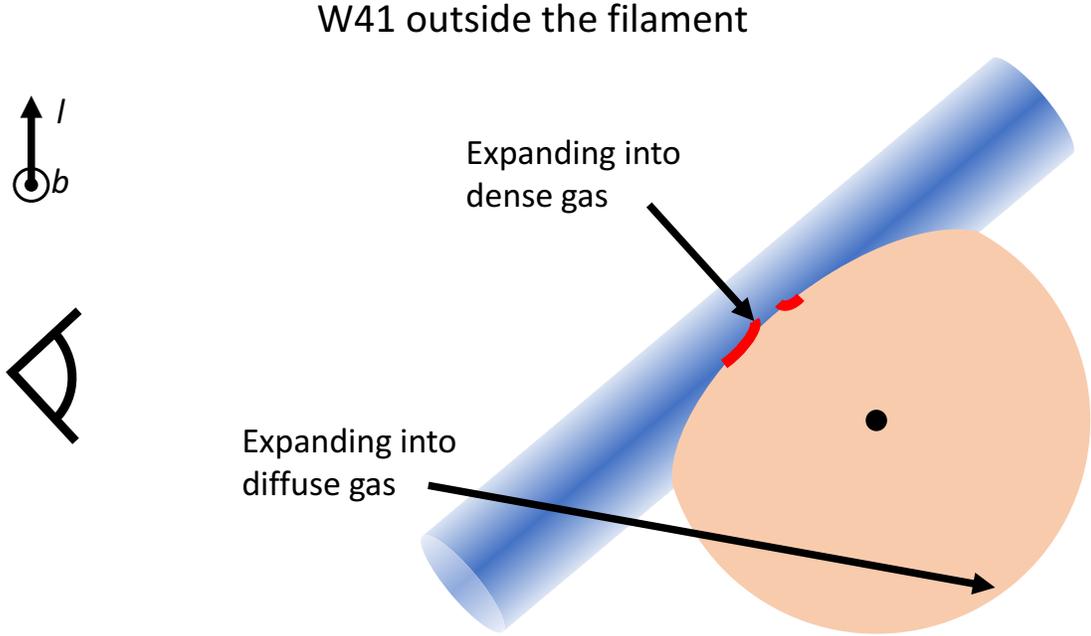}
\caption{Illustration of top-down view of the scenario in which W41 formed behind GMC G23.0-0.4. The meaning of the shapes is given in the caption of Figure~\ref{fig:cartoon1}.}
\label{fig:cartoon2}
\end{figure}

Considering that there are multiple background sources that confuse the interpretation of any potential redshifted gas, the gas kinematics in G23.33-0.30 may offer additional insight into the interaction geometry. Given that G23.33-0.30 exhibits blueshifted emission as a result of its interaction with W41, W41's progenitor must be behind G23.33-0.30. The details of this interaction also depend on the positions of G23.33-0.30 and W41's progenitor relative to G23.0-0.4. Considering G23.33-0.30's large column density, its apparent position along the spine of the G23.0-0.4, and the necessarily high-mass nature of W41's progenitor, it is plausible that both G23.33-0.30 and W41's progenitor formed near the dense central regions of their natal MCs. If this assumption is valid, then at least two viable scenarios that could explain the geometry: either W41's progenitor formed within the GMC and the shock front is perpendicular to the long axis of the filament, or W41's progenitor formed within a background MC and its shock front is roughly parallel with the long axis of the filament. These two scenarios are illustrated in Figures~\ref{fig:cartoon1} and \ref{fig:cartoon2}. Unfortunately, neither of these explanations account for the fact that the eastern edge of the filament is more blueshifted than the western edge. If G23.33-0.30 is closer to the edge of the GMC than assumed, the shock could sweep past the bottom edge of the filament more quickly due to the decreased density on the outskirts of the GMC. Another alternative is that projection affects are confusing our interpretation of the filament's velocity structure. Regardless of the uncertain impact geometry, W41's association with G23.33-0.30's blueshifted gas component and large velocity dispersions is well founded. 

\subsection{Possible Negative Feedback from SNR W41} 

Although shocks from supernovae may drive much of the turbulence in the ISM \citep{2016ApJ...822...11P}, the influence of these shocks on the star formation process is an open question. Simulations of shock-cloud interactions that include dense substructures demonstrate that the more diffuse molecular gas is efficiently stripped from MCs, while such an impact forms a bow shock around a sufficiently dense molecular core \citep{2005ApJ...633..240P}. 
It is possible that the emission deficits in Figure~\ref{fig:dec-vlsr} at lower $V_{\rm{LSR}}$ represent a wake formed behind the cores, which would suggest that some dense gas associated with the cores will remain relatively unperturbed by the passage of the shock. On the other hand, the virial parameters derived from the $^{13}$CO emission suggest that the cores are too turbulent to collapse. Furthermore, given that the molecular cores are associated with the intermediately shocked component, they may not yet have experienced as significant an impulse from the shock as compared to the northern section of G23.33-0.30. Indeed, MM3, the core furthest from the turbulent component, has the largest $\frac{\rm{C}^{18}\rm{O}(2-1)}{^{13}\rm{CO}(2-1)}$ brightness ratio. This indicates that MM3 has the highest CO(2-1) optical depth, implying that it is also the densest of the cores. If W41's shock passes through the filament and continues to interact with the molecular cores, their velocity dispersions may increase, potentially resulting in the dispersal of the cores. Although we cannot determine whether W41's shock is responsible for the turbulent velocity dispersions of the cores, our data are consistent with negative feedback from the SNR.

\section{Conclusion}
\label{sec:con}

G23.33-0.30 is a massive IRDC filament that exhibits broad molecular line widths and narrow NH$_3$(3,3) line emission. We have imaged the filament using the VLA (NH$_3$(1,1) through (6,6)), the SMA ($^{13}$CO, C$^{18}$O, 1.3 mm continuum), and the ACA (SiO(5-4), CS(5-4), 1.3 mm continuum) and we have drawn the following conclusions from our data:

\begin{enumerate}
\item We have confirmed the nonthermal nature of three NH$_3$(3,3) masers that peak near $V_{\rm{LSR}} = 56$ km s$^{-1}$ and discovered a fourth NH$_3$(3,3) maser at $V_{\rm{LSR}} = 76$ km s$^{-1}$. 
\item The ACA observations revealed broad SiO(5-4) emission throughout the filament, indicating the presence of a highly turbulent and extended shocked gas component. Because protostellar outflows cannot reproduce the observed SiO emission, high-mass protostellar outflows do not produce G23.33-0.30's NH$_3$(3,3) maser emission. The widespread nature of the SiO emission indicates a shock acting on larger scales. 
\item The NH$_3$ emission displays a velocity gradient along the length of the filament, with a significant portion of the filament apparently blueshifted by $\sim 10-17$ km s$^{-1}$ with respect to the rest of the filament. G23.33-0.30 also exhibits a velocity discontinuity across the width of the filament, which separates the shocked, turbulent component from the pre-shock component. 
\item Our LTE NH$_3$ model fitting to the NH$_3$(1,1), (2,2), (4,4), and (5,5) data has provided maps of the NH$_3$ rotational temperature ($T_{\rm{rot}}$), velocity dispersion ($\sigma$), and LSR velocity ($V_{\rm{LSR}}$), which show that the shocked component is hotter, more turbulent, and blueshifted compared to the pre-shock component. The correlation among $T_{\rm{rot}}$, $\sigma$, and $V_{\rm{LSR}}$ implies that the shock is simultaneously accelerating, heating, and injecting turbulent energy into the shocked filament gas. In addition, the increased ortho-NH$_3$ abundance inferred from the large $\frac{\Delta T_B(6,6)}{\Delta T_B(5,5)}$ and $\frac{\Delta T_B(6,6)}{\Delta T_B(4,4)}$ brightness temperature ratios imply dust heating.
\item G23.33-0.30 resides within the GMC G23.0-0.4, which previous authors have speculated is undergoing a SNR-MC interaction with W41. Our interferometric data and the GRS $^{13}$CO data provide additional evidence of this interaction, which suggests that W41's shock is the common cause for the observed gas kinematics on large and small scales. 
\item Although W41's impact geometry remains ambiguous, its interaction with G23.33-0.30 and its plane-of-sky diameter imply that it is at a distance of $d\leq20$ pc from G23.33-0.30.  
\item The SMA 1.3 mm continuum data revealed dust cores embedded within G23.33-0.30. Although G23.33-0.30 appears to have sufficient mass ($M \sim 600$~M$_{\odot}$) to form a high-mass star, the observed gas kinematics suggest that the filament is presently being displaced, and potentially dispersed, by the SNR shock. Likewise, our virial analysis ($\alpha = 4-9$) suggests that the cores are also unlikely to collapse. Thus, our data are consistent with negative feedback from the SNR.
\end{enumerate}

\section*{Acknowledgments} 
We thank the anonymous referee for useful comments which improved the paper. This research was supported by the National Science Foundation Grant AST-1616635. The National Radio Astronomy Observatory is a facility of the National Science Foundation operated under cooperative agreement by Associated Universities, Inc. This paper makes use of the following ALMA data: ADS/JAO.ALMA$\#$2016.1.01345.S. ALMA is a partnership of ESO (representing its member states), NSF (USA) and NINS (Japan), together with NRC (Canada), MOST and ASIAA (Taiwan), and KASI (Republic of Korea), in cooperation with the Republic of Chile. The Joint ALMA Observatory is operated by ESO, AUI/NRAO and NAOJ. The Submillimeter Array is a joint project between the Smithsonian Astrophysical Observatory and the Academia Sinica Institute of Astronomy and Astrophysics and is funded by the Smithsonian Institution and the Academia Sinica. This research has made use of NASA's Astrophysics Data System.
\software{PySpecKit (Ginsburg et al. http://doi.org/10.5281/zenodo.12490), APLpy (Robitaille and Bressert, 2012)}

\bibliography{G23_Paper_arxiv}
\end{document}